\documentclass[12pt,preprint]{aastex}
\newcommand{\ie}{{\em i.e.}}
\newcommand{\uHz}{\mbox{$\mu$Hz}}

\shorttitle{Mode Fitting for Very Long Time Series}
\shortauthors{Korzennik}      
\received{}
\title{A Mode Fitting Methodology\\ Optimized for Very Long Time Series}
\author{S.~G.~Korzennik
}
\affil{Harvard-Smithsonian Center for Astrophysics, Cambridge, MA 02138, USA}
\begin{document}
\begin{abstract}
I describe and present the results of a newly developed fitting methodology
optimized for very long time series. The development of this new methodology
was motivated by the fact that we now have more than half a decade of nearly
uninterrupted observations by GONG and MDI, with fill factors as high as
89.8\% and 82.2\% respectively.  It was recently prompted by the availability
of a 2088-day-long time series of spherical harmonic coefficients produced by
the MDI team.
The fitting procedure uses an optimal sine-multi-taper spectral estimator --
whith the number of tapers based on the mode linewidth, the complete leakage
matrix (\ie, horizontal as well as vertical components), and an asymmetric
mode profile to fit simultaneously all the azimuthal orders with individually
parametrized profiles.
This method was applied to 2088-day-long time series of MDI and GONG
observations, as well as 728-day-long subsets, and for spherical harmonic
degrees between 1 and 25. The values resulting from these fits are
inter-compared (MDI versus GONG) and compared to equivalent estimates from the
MDI team and the GONG project. I also compare the results from fitting the
728-day-long subsets to the result of the 2088-day-long time series. This
comparison shows the well known change of frequencies with solar activity --
and how it scales with a nearly constant pattern in frequency and
$m/\ell$. This comparison also shows some changes in the mode linewidth and
the constancy of the mode asymmetry.
\end{abstract}
\section{Motivation}

  Several methodologies for the precise measurements of p-mode frequencies
have been developed over the years \citep[see, amongst other,][]
{Libbrecht:88,
 AndersonEtAl:90,
 Korzennik:90,
 Schou:92,
 DuvallEtAl:93,
 ToutainEtAl:98,
 AppourchauxEtAl:98a,
 Rabello-Soares+Appourchaux:99,
 Thiery:00a, 
 Jimenez-Reyes:01}. These techniques have evolved as the quality of the
observations have improved over the years.  More recently, such methods have
produced frequency and rotational frequency splitting tables -- typically
based on 36-day-long and 72-day-long time series for GONG and MDI data
respectively -- that have been the basis for countless inferences of the
structure and dynamics of the solar interior. 

  Some eight years after the deployment of the GONG network and the launch of
the SOHO spacecraft, we now have access to more than half a decade of nearly
uninterrupted times series of solar observations. The MDI team has recently
produced a 2088-day-long time series of spherical harmonic coefficients at a
one minute sampling interval, namely a time series in excess of 3 millions
points. For low-degree and low-frequency modes the availability of such a very
long time series provides a unique opportunity to develop a mode fitting
methodology that exploits the properties of such an exquisite data set.

  I present here a new methodology that I have developed to fit modes using
very long time series. This methodology includes an optimal multi-taper
spectral estimator, the complete leakage matrix (\ie, horizontal as well as
vertical components), an asymmetric profile and the simultaneous fitting of
individual profiles at all the azimuthal orders ($m$) for a given ($n$,
$\ell$) mode. The contamination by nearby modes ($n'$, $\ell'$) within the
fitting range is also included. Since simultaneous fitting on these
contaminants is impractical, the fitting procedure is iterated -- \ie, the
characteristics of the contaminants used in the fitting correspond to the
values fitted at the previous iteration.

  The primary goal for this work was to extend the mode fitting to low-order
and low-degree modes. The low degree modes carry information on the structure
and dynamics of the deep interior while the low-order low-degree modes are
long lived (\ie, show very narrow peaks) and can thus be measured with very
high precision. Since their amplitude is small, they will emerge above the
background noise only for very long time series. The secondary goals were to
fit simultaneously all the modes instead of using the traditional polynomial
expansion in ${m}/{\ell}$, while using the complete leakage matrix, as well as
to include the well known mode profile asymmetry.

  I describe the data sets I used in Section 2, and present the details of the
methodology in Section 3. In Section 4 I present the results of fitting both
MDI and GONG 2088-day-long co-eval time series. These results are compared to
equivalent estimates from the MDI team and the GONG project. I also present
the results of fitting two-year-long subsets of that time series for both data
sets, and how the fitted parameters change with epoch and henceforth with
solar activity.

\section{Data Sets Used}

  The work presented here is based on times series of spherical harmonic
coefficients computed from full-disk observations by the MDI and GONG
instruments and limited to $\ell \le 25$. The MDI 2088-day-long time series
was used to develop the methodology. It was then applied to the co-eval GONG
data set. Both data sets were also subdivided in five 728-day-long overlapping
segments (actually 1,048,576 minute long, \ie: $1024^2$), each offset by some
364 days (\ie, 524,288 minutes) from the previous one. The last 728-day-long
segment was completed by augmenting the 2088-day-long time series by an extra
97 days (\ie, 139,008 minutes) to fill it completely. The respective ranges of
all the time series analyzed are shows in Table~\ref{table:TimeRanges}.

\begin{table}[t]
\begin{tabular}{l|c|cc|}
Length & Segment No & From & To \\ \hline
2088-day & n/a & 04/30/1996 at 23:59:30 UT & 01/17/2002 at 23:59:26 UT \\
728-day  &  1  & 04/30/1996 at 23:59:30 UT & 04/29/1998 at 04:14:30 UT \\
728-day  &  2  & 04/30/1997 at 02:07:30 UT & 04/28/1999 at 06:22:30 UT \\
728-day  &  3  & 04/29/1998 at 04:15:30 UT & 04/26/2000 at 08:30:30 UT \\
728-day  &  4  & 04/28/1999 at 06:23:30 UT & 04/25/2001 at 10:38:30 UT \\
728-day  &  5  & 04/26/2000 at 08:31:30 UT & 04/24/2002 at 12:46:30 UT \\
\end{tabular}
\caption{Respective time ranges for the time-series analyzed
\label{table:TimeRanges}}
\end{table}

The fill factors of the 2088-day-long time series, before detrending, are
89.8\% and 82.2\% for MDI and GONG observations respectively. The time series
of spherical harmonic coefficients were detrended using a 21-minute-long
running mean and clipped with a 3$\sigma$ rejection threshold. The rejection
reduced the fill factor by less than one percent for the MDI time series and
by less than two percents for the GONG time series.  For the MDI observations,
discontinuities resulting from instrumental reconfigurations were removed
by detrending over sub-intervals delineated by these reconfigurations.

\section{Methodology}

  The key original aspects of this new methodology are the use of an {\em
optimal} multi-tapered spectral estimator, the simultaneous fitting of all $m$
spectra and the use of an asymmetric profile. The fitted model includes both
radial and horizontal leakage components and the leakage of nearby modes
inside the fitting range.

\subsection{Spectral Estimator}

  The $N^{\rm th}$ order sine multi-taper, defined as
\begin{equation}
   P_{\ell, m}^{(N)}(\nu) 
     = \sum_{k=1}^{N} 
  \left|{\rm FFT}\left[
   \sin(\frac{\pi\,k\,i}{M+1}) c_{\ell, m}(t_i)
  \right]\right|^2
\end{equation}
was used as power spectrum estimator, where $c_{\ell, m}(t_i)$ represents the
spherical harmonic coefficient for $\ell$ and $m$ at the time $t_i$, and $M$
is the length of the time series. Sine multi-tapered power spectra were
computed with an oversampling factor of 2, and for a pre-selected list of
number of tapers. For the 2088-day-long time series, I used 5, 9, 21, 45 and
91 tapers, while for the 728-day-long time series I used 3, 7, 15, 31 and 63
tapers. The choice of the optimal number of tapers amongst that list is
explained below.

\subsection{Mode Fitting}

The fitting procedure uses a downhill simplex minimization
\citep{Nelder+Mead:65, PressEtAl:92} to fit simultaneously, and in the
least-squares senses, all the multiplets for a given mode, \ie\ all $m$ for a
given $n, \ell$. The fitting is done iteratively, over a frequency range
limited to only encompass the closest spatial leaks ($\delta m = \pm 2$,
$\delta \ell = 0$), using the optimal multi-taper power spectrum and fitting
only for the modes whose amplitudes are above some prescribed threshold.

The fitted profile is an asymmetric Lorentzian:
\begin{equation}
P_{n, \ell,m}(\nu) = 
 \frac{1+\alpha_{n,\ell} (x_{n,\ell,m}-\alpha_{n,\ell}/2)}{x_{n,\ell,m}^2+1}
\end{equation}
where
\begin{equation}
 x_{n,\ell,m} = \frac{\nu - \nu_{n,\ell,m}}{\Gamma_{n,\ell}/2}
\end{equation}
The mode linewidth, $\Gamma_{n,\ell}$, and its asymmetry, $\alpha_{n,\ell}$,
are assumed to be independent of the azimuthal order, $m$. The power spectrum
is thus modeled as the superposition of the mode profile and the spatial leaks
present in the fitting range:
\begin{eqnarray}
P_{\ell, m}(\nu) = 
  &  &A_{n,\ell,m} P_{n, \ell, m}(\nu) + B_{n, \ell, m} \\\nonumber
  &+&  \sum_{m'}
       A_{n,\ell,m'}\, C(n, \ell, m'; n, \ell, m)\, P_{n, \ell, m'}(\nu) 
       \\\nonumber
  &+&   \sum_{n', \ell', m'} 
       A_{n',\ell',m'}\, C(n', \ell', m'; n, \ell, m)\, P_{n', \ell', m'}(\nu)
\end{eqnarray}
where $A$ represents the respective mode power amplitudes, $B$ the noise
background levels and $C$ the leakage coefficients. For obvious procedural
reasons, I will refer to the terms in the first summation as spatial leaks
(same $n$ and $\ell$) and the ones in the second summation as mode
contamination (leaks from a different $n$ and $\ell$). The sum on $m'$ is
actually limited to $m'=m-2$ and $m'=m+2$ by the choice of the fitting range
(see below). The sum on $n',\ell',m'$ is included only if a nearby mode
($|\ell-\ell'| \le 3$ \& $|m-m'| \le 3$) fall within the widened fitting
window (\ie, the fitting range expanded by 40\% to include the tail of nearby
leaks).

  In the absence of mode contamination in the fitting range, there are
2$\ell$+1 amplitude, frequency, and background parameters plus one linewidth
and asymmetry parameters, or $3\times(2\ell+1)+2$ coefficients to fit using
$2\ell+1$ sections of spectra.  By performing a simultaneous fit, the main
peak as well as the spatial leaks are used to constrain the mode parameters,
under the implicit assumption that the leakage coefficients are perfectly well
known.

  For $\ell \le 25$, the mode contamination is dominated by $n-n' = \pm 1$ and
$\ell-\ell' =\mp 3$, and is almost confined to the high frequency modes.  When
present, this mode contamination was added in the fitted model by using the
mode parameters resulting from the previous iteration, and leaving them fixed.
The fitting procedure is therefore repeated --- \ie\ iterated --- until the
mean change in the mode frequencies between iterations drops below some
prescribed threshold.

\subsubsection{Fitting Range}

  The fitting frequency range is set to be $\tilde\nu_{n,\ell,m} \pm
\delta\nu$, where $\tilde\nu_{n,\ell,m}$ represents some estimate of the
multiplet frequency at the previous fitting step (see below) and $\delta\nu$
is given by
\begin{equation}
\delta\nu = 4\, \Gamma^{(\rm eff)}_{n, \ell} + \Delta\nu
\end{equation}
where $\Gamma^{(\rm eff)}$ is the effective mode linewidth and $\Delta\nu=800$
nHz. The factor 4 ensures a good sampling of the mode profile while the
additional 800 nHz is added to always include the $\delta m = \pm 2$, $\delta
\ell = 0$ spatial leaks. This range is set to never be smaller than 1.2 \uHz\
at low frequencies and reduced, at high frequencies, to not exceed the
mid-point from the frequency of the previous order and to the next order. The
effective mode linewidth, $\Gamma^{(\rm eff)}_{n, \ell}$, is estimated by
\begin{equation}
(\Gamma^{(\rm eff)}_{n, \ell})^2 = \tilde\Gamma^2_{n, \ell} + \Gamma_{r, N}^2
\end{equation}
where $ \tilde\Gamma_{n, \ell}$ is some estimate of the mode linewidth (see
below) and $\Gamma_{r, N}$ is the resolution of the $N^{\rm th}$ order
multi-taper power spectrum, given by
\begin{equation}
\Gamma_{r,N} = N\ \Gamma_r = \frac{N}{T}
\end{equation}
and where $T$ is the time interval span by the time series.

\subsubsection{Optimal Multi-Taper}

The optimal $N^{\rm th}$ order multi-tapered power spectrum is defined as the
highest order multi-taper spectrum from a pre-selected list having a
resolution at least five times better than the effective mode linewidth,
whenever possible. The value of $N$ is thus selected to satisfy:
\begin{equation}
 \frac{\tilde\Gamma_{n,\ell}}{5} \ge \Gamma_{r,N} = N\,\Gamma_r
\end{equation}
Mainly for convenience, an estimate of the mode linewidth was used for this
selection and for the determination of the mode fitting range. This estimate
was computed from a polynomial parameterization as a function of frequency,
based on previous published estimates \citep{Schou:99}. The logarithm of the
mode linewidth was estimated by a 6th order polynomial in $\nu$.

\begin{figure}[!p]
\includegraphics{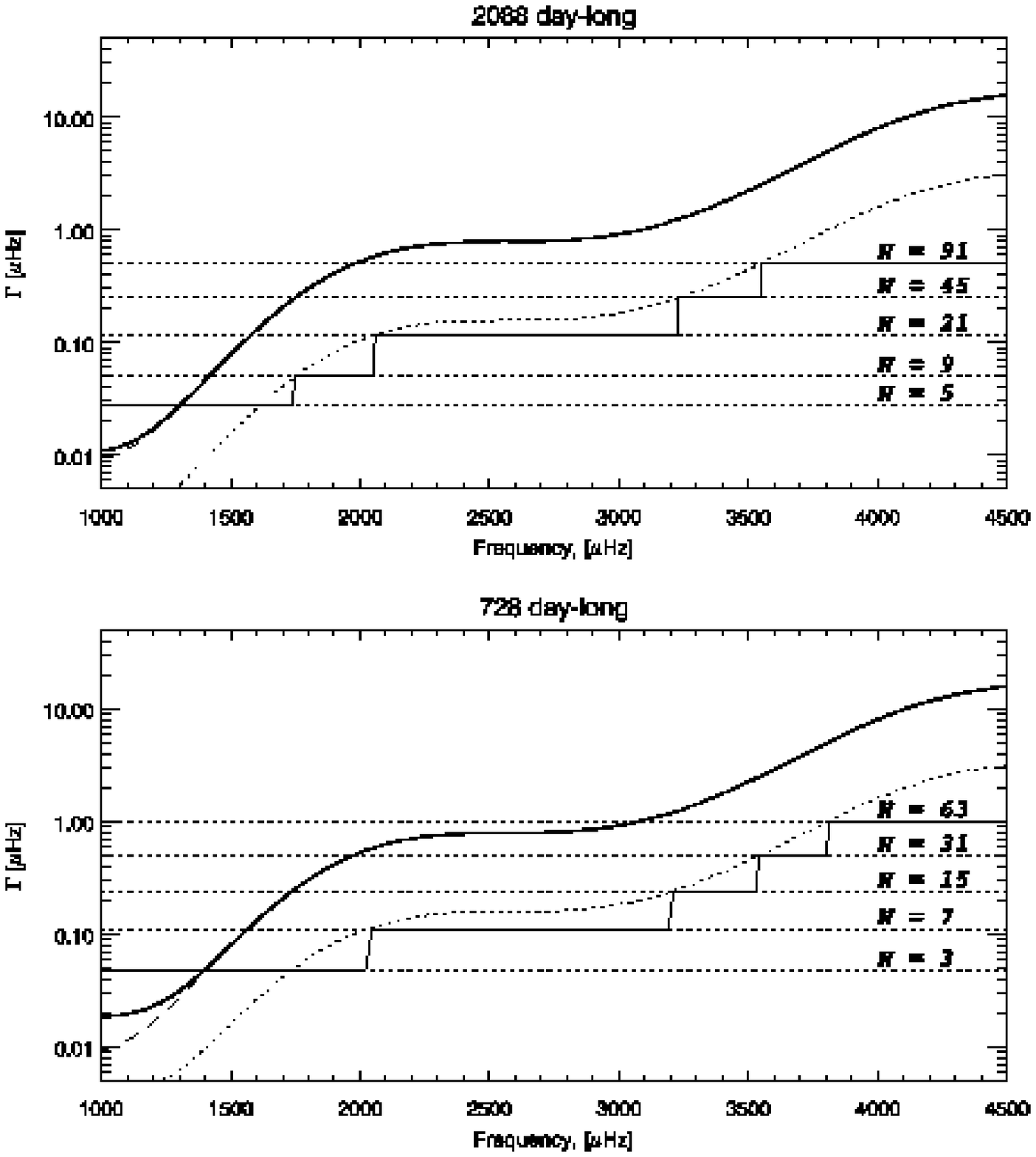}
\caption{Mode linewidth (long dash) and effective linewidth (solid bold
curve), compared to multi-tapers spectral resolution (dotted lines). The
optimal number of multi-tapers (see description in text) is indicated by the
stepwise solid line, and is such that the spectral resolution remains 5 times
smaller than the effective linewidth, whenever possible. Top and bottom panels
correspond to 2088-day-long and 768-day-long time series, respectively.
\label{figure:whichmt}}
\end{figure}

   Figure~\ref{figure:whichmt} compares the estimate of the mode linewidth to
the $N^{\rm th}$ order multi-tapered spectrum resolution, for the pre-selected
list of values of $N$, and indicates the number of tapers selected for the
fitting.

\subsubsection{Detectability Threshold}

   At low frequencies, the mode amplitude becomes comparable if not smaller
than the background noise all the while the mode linewidth becomes smaller
than the spectral resolution, forcing the fitting procedure to hunt for small
and narrow peaks. This can easily lead to confusing a noise spike with a mode
peak and can lead to what is sometimes referred as {\em fitting the grass}.
To prevent such ``grass fitting'', only modes with a power amplitude 3 times
greater than the root-mean-squares (RMS) of the residuals to the fit were
kept. As the fitting proceeds a sanity check rejects any mode whose amplitude
has dropped below that threshold. If this happens, the amplitude of the mode
is set to zero and its amplitude and frequency are no longer fitted, leaving
the fitting procedure to only adjust the background term to the noise.

  As a result of the line of sight apodization of any radial velocity
observation, the near sectoral modes ($m \approx \pm\ell$) emerge above the
noise background before the near zonal ones ($m \approx 0$). By not using a
polynomial expansion in $m$, I end up fitting only the azimuthal degrees that
are large enough, and I avoid any potential bias in the estimate of the
frequency splittings based on an expansion that would be unevenly constrained.

\subsubsection{Fitting Procedure \& Initial values}

  At each iteration the fitting is done in steps, \ie\ not all the parameters
are adjusted simultaneously right away. The whole procedure is iterated to
include progressively better and better estimates of the mode contamination.

  An initial value of all the mode parameters is first needed. The initial
guess for the mode frequency was computed by using previous estimates of the
singlet frequencies ($\nu_{\ell,m}$) and of the polynomial expansion
coefficients for the frequency splittings. The frequency table was extended at
low frequencies by looking at ``derotated'' $m$-averaged spectra while the
frequency splittings table was extended with values that produced nearly
``straight'' derotated $m$-averaged spectra. The initial values for the
linewidth and the asymmetry were computed from a polynomial parameterization
in term of frequency, also based on previous estimates, and extrapolated when
needed. The initial amplitude and background parameters were computed from
averaging appropriate portions of the section of the spectrum being fitted and
then performing a polynomial fit of these values as a function of $m$ with a
3$\sigma$ rejection of outliers.

  To make sure that the result is not biased by the initial guess of the
parameters, the fitting steps at the very first iteration were different from
the ones taken during the consecutive iterations. The first iteration included
additional steps where the fitting was done on individual spectra (\ie, one
$m$ at a time) instead of the simultaneous fitting of all the $m$.

  A sanity check is performed after each step to remove from the fitting any
peak with too low of an amplitude. The amplitude of such peaks are set to zero
and their amplitude and frequency is no longer fitted.  This check is followed
by a readjustment of the fitting interval by using for the center of the
interval smoothed values resulting from a polynomial fit of the frequencies as
a function of $m$, with a 3$\sigma$ rejection of outliers.

\subsubsection{Leakage Coefficients}

  The first leakage coefficients I used for the MDI observations were based on
the direct computation of the spatial leakage from the leakage equations (see
Section 3.1 of \cite{KorzennikEtAl:2004}). Such calculation leads to an
approximation of the actual leakage coefficients as it remains oversimplified
by ignoring effects like the detector finite pixel size and the on-board
Gaussian weighting performed on the MDI {\em Structure Program} images.

  A more sophisticated leakage matrices computation was carried out by
\citet{Schou:99}, by constructing simulated images corresponding to the
line-of-sight contribution of each component of a single spherical harmonic
mode. These images were then decomposed into spherical harmonic coefficients
using the same numerical decomposition used to process the observations. This
approach allows to include the effect of the finite pixel size of the detector
and incorporates the above mentioned Gaussian weighting. It also takes into
account {\em de facto} the effects of both the foreshortening at high
lattitudes and the image apodization performed in the spatial decomposition.

  In both cases, the horizontal and vertical components were computed and the
horizontal to vertical displacement ratio, $\beta$, taken to be the
theoretical prediction. Using a simple outer boundary condition, \ie\ that the
Lagrangian pressure perturbation vanishes ($\delta p = 0$), the small
amplitude oscillation equations for the adiabatic and non-magnetic case lead
to an estimate of the ratio $\beta$, given by: 
\begin{equation}
  \beta_{n, \ell} = \frac{G\;M_{\odot}\;L}{R_{\odot}^3\; \omega^2_{n, \ell}} =
                  \frac{\nu_{0, \ell}^2}{\nu_{n, \ell}^2}
\label{eq:beta}
\end{equation}
where $G$ is the gravitational constant, $M_{\odot}$ is the solar mass,
$R_{\odot}$ is the solar radius, $\omega$ is the cyclic frequency
($\omega=2\pi\nu$) and $L^2 = {\ell(\ell+1)}$. 

  Leakage matrices were also computed by Schou (private communication, 2004)
specifically for the GONG observations. The GONG project has also recently
computed leakage matrices (Howe \& Hill, private communication, 2004) for the
GONG observations, but have only included the vertical component. I have
therefore used for the fitting of the GONG data Schou's complete leakage
matrix.  In all cases, since the time series are long compared to a year I
used leakage matrices computed for $B_{\rm o}=0$.

\begin{figure}[!p]
\includegraphics{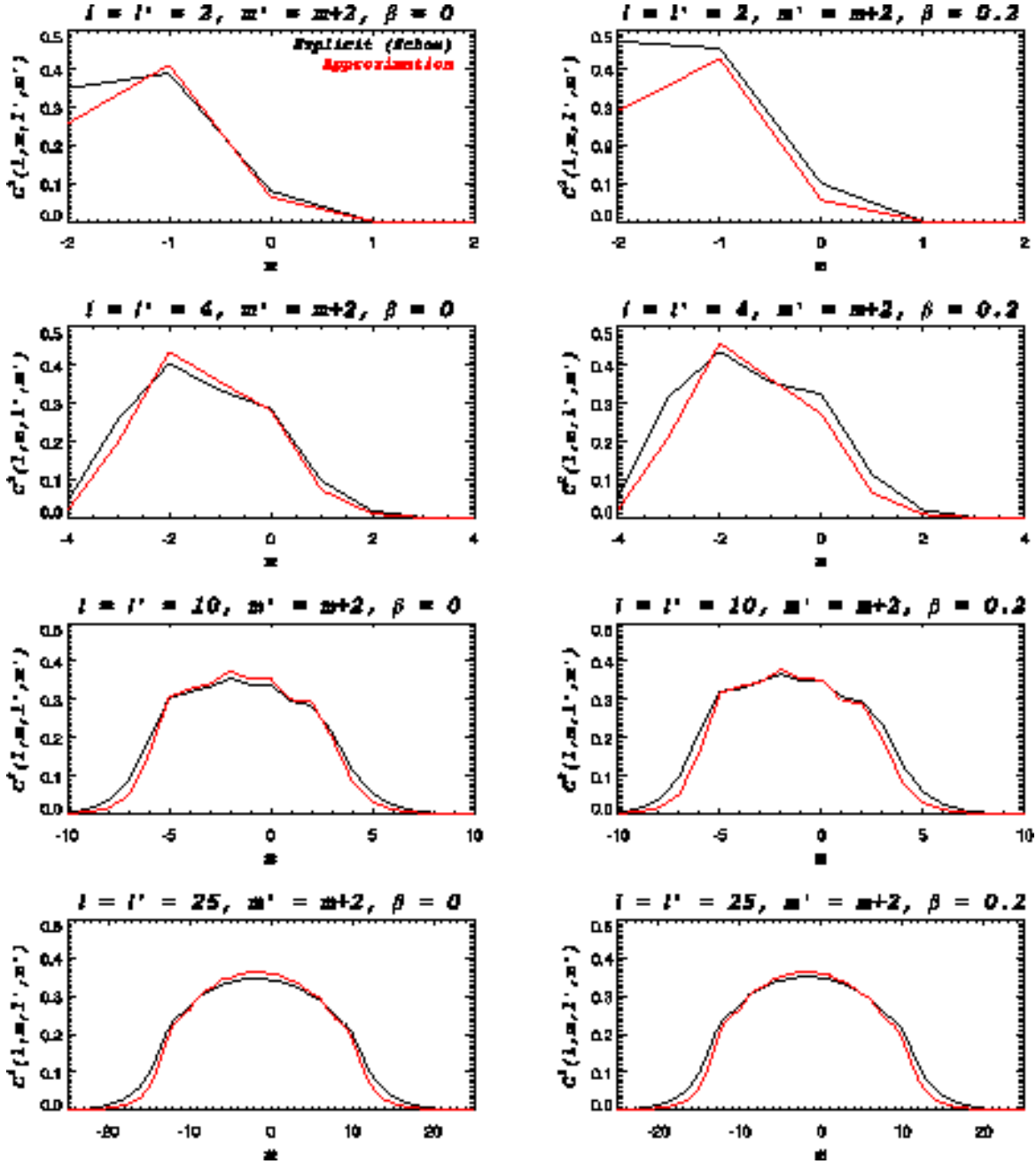}
\caption{Comparison of leakage matrix, for selected values of $\ell$ and
$\beta$, resulting from a direct computation (analytical approximation) to the
values computed explicitly by Schou for the MDI instrument. The RMS of the
differences is around 4 to 7\% below $\ell=10$ and less than 2\% above. The
``wider'' leakage seen in the explicit computation is primarly the result
of including the Gaussian weighting of the MDI images for the {\em Structure
Program} data.
\label{fig:compare_leaks}}
\end{figure}

\begin{figure}[!p]
\includegraphics{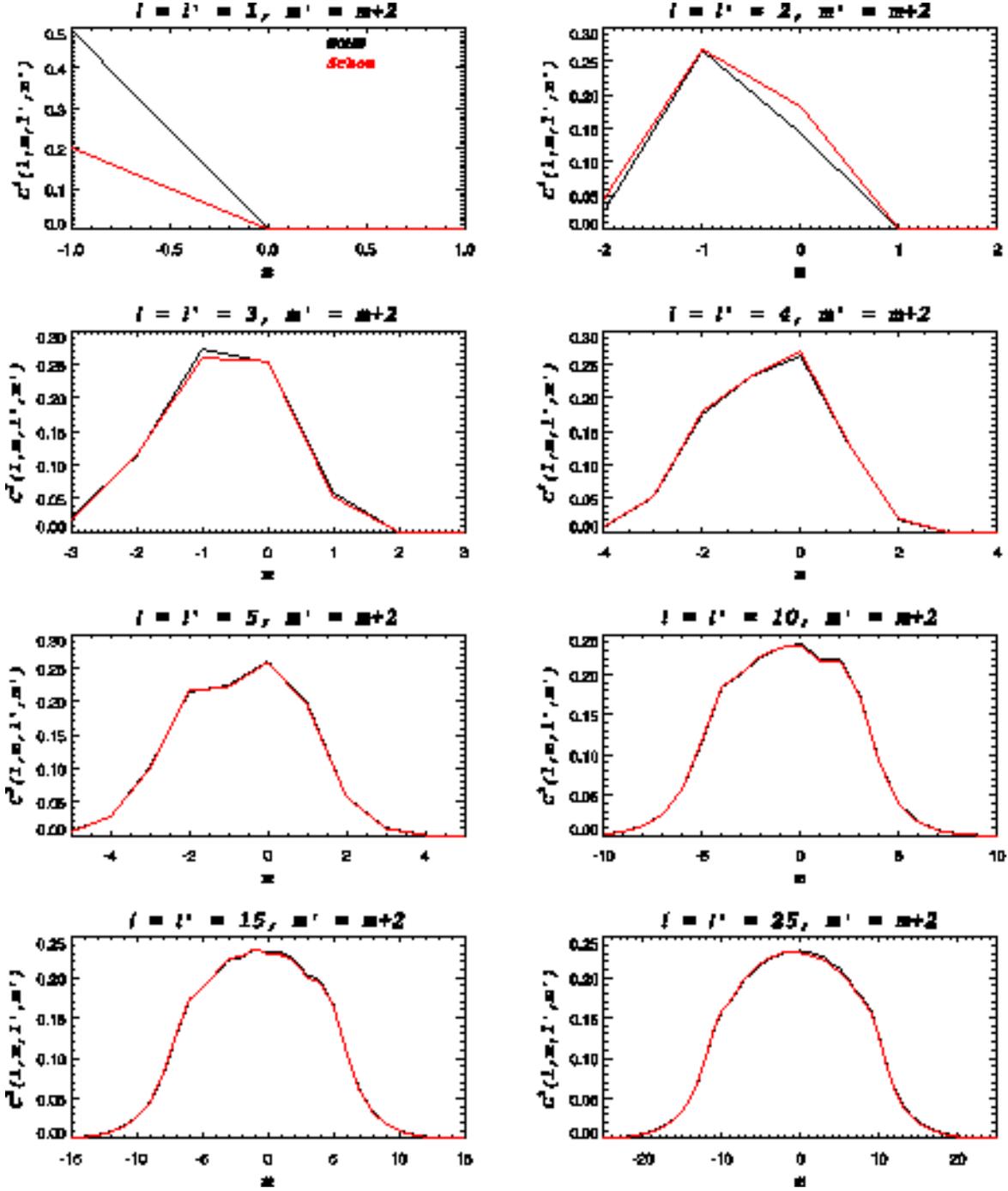}
\caption{Comparison of the radial component of the leakage matrix computed for
the GONG observations by the GONG project (black) and by Schou (red). The RMS
of the differences is less than 0.2\% above $\ell=4$.
 \label{fig:compare_leaks_gong}}
\end{figure}

  Figure~\ref{fig:compare_leaks} compares the leakage matrix, for selected
values of $\ell$ and $\beta$, resulting from my direct and approximate
computation to the values computed explicitly by Schou for the MDI
instrument. Figure~\ref{fig:compare_leaks_gong} compares the radial component
of the leakage matrix as computed by the GONG project and by Schou, for the
GONG observations. The differences seen in Figure~\ref{fig:compare_leaks}
should not come as a surprise: the wider leakage seen in the explicit
computations is primarly the result of the Gaussian weighting.  The RMS of the
differences for the radial component of the GONG leakage matrices, as shown in
Figure~\ref{fig:compare_leaks_gong}, is less than 0.2\% above $\ell=4$. The
discrepancy at $\ell=1$ is due to the fact that the GONG values includes
already the effect of subtracting the image average.

\subsubsection{Error Bars Computation}

  Uncertainties on all the fitted parameters were estimated from the
covariance matrix of the problem. This covariance matrix was estimated from
the Hessian matrix \citep{PressEtAl:92} using numerical estimates of the
second derivative of the merit function. The increments appropriate for the
estimate of these derivatives were based on the size of the shrunk simplex
resulting from the fitting.

\section{Results}

\begin{figure}[!p]
\includegraphics[scale=.9]{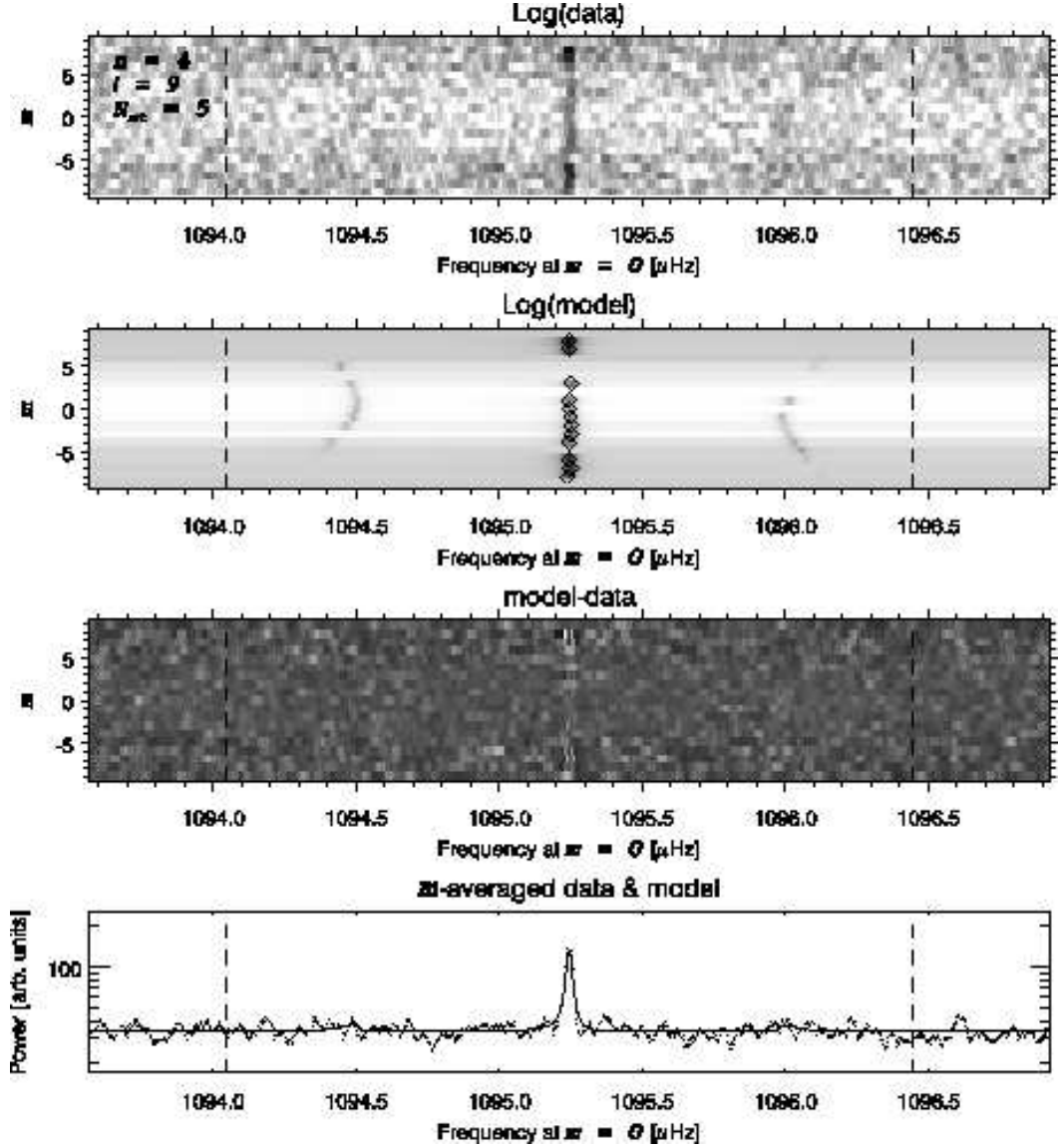}
\caption{Example of fitting, for MDI 2088-day-long time series, for $\ell=9$,
$n=4$ and $N=5$. The top panel shows a fraction of the derotated power
spectra (5th sine multi-taper). The second panel from the top shows the model
of the power spectra, diamonds indicate the location of the mode
frequencies. The third panel from the top shows the residuals and the bottom
panel show $m$-averaged profiles (data and model). The vertical dash lines
delineates the frequency range used for the fitting. Notice how for this low
order case some of the modes amplitudes were not large enough to be fitted.
\label{fig:example1}}
\end{figure}

\begin{figure}[!p]
\includegraphics{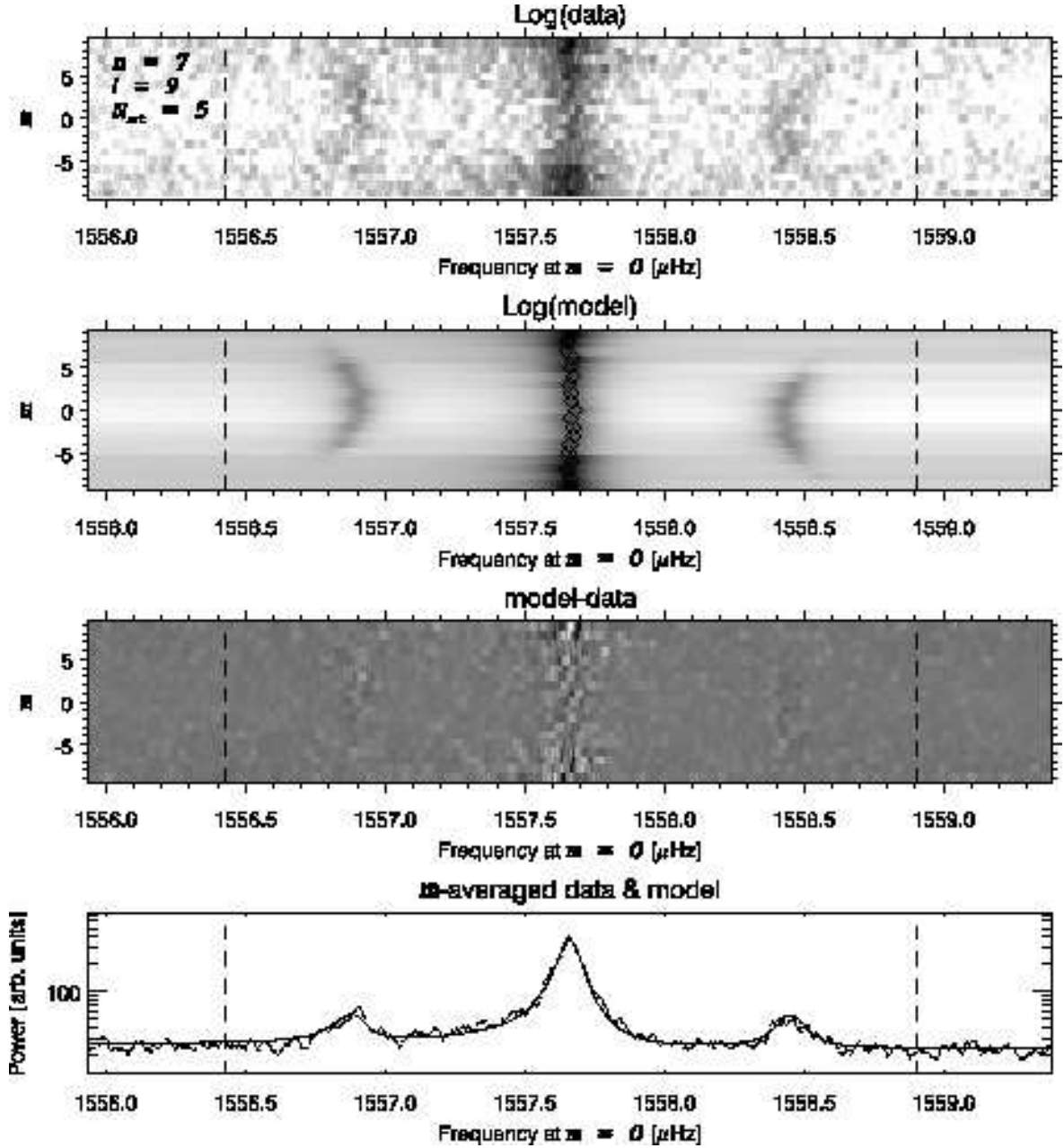}
\caption{Example of fitting, as in Figure~\ref{fig:example1}, but for
$n=7$. In this case the mode and the leaks are still well resolved and all the
modes are well above the background.
         \label{fig:example2}}
\end{figure}

\begin{figure}[!p]
\includegraphics{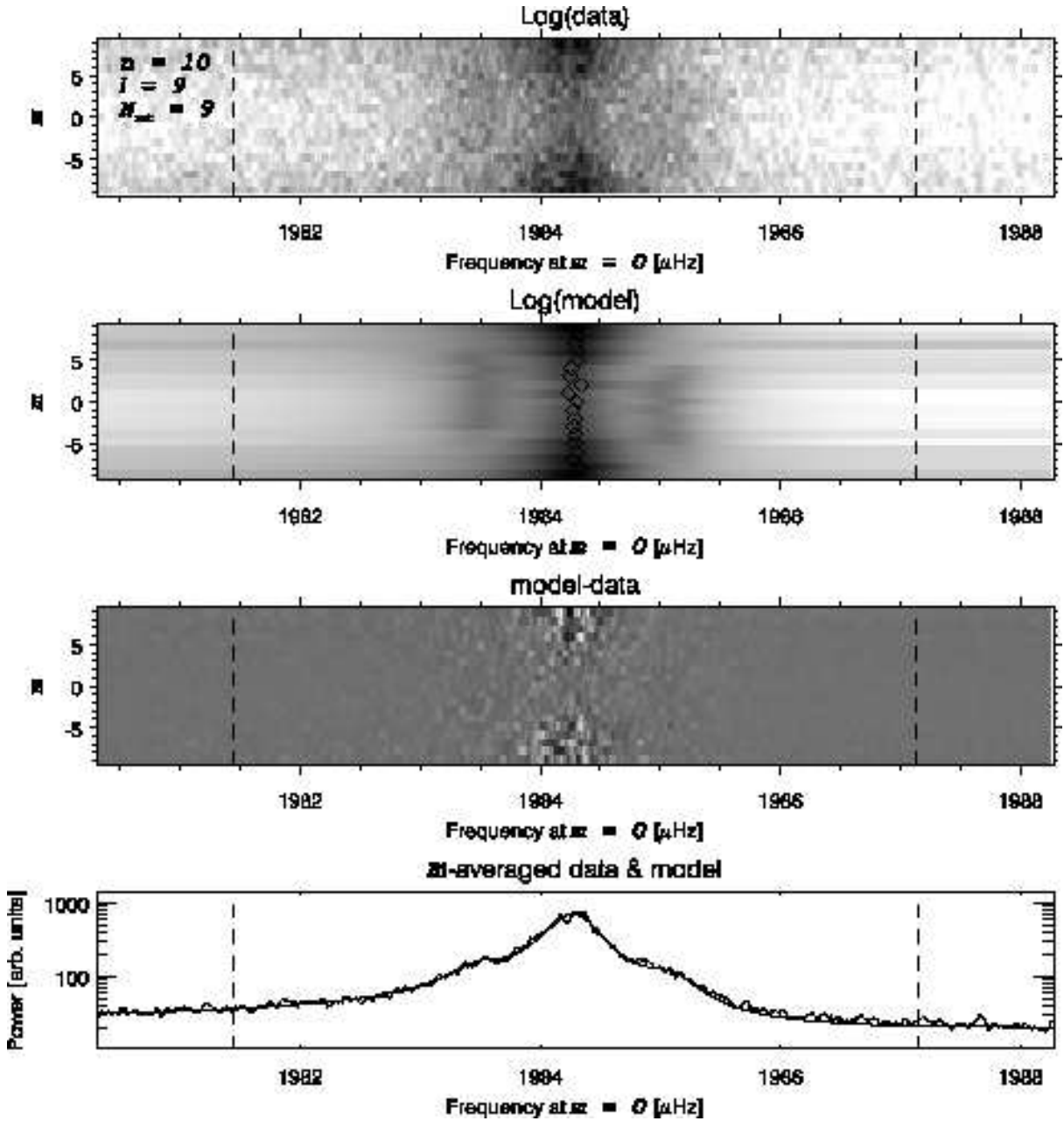}
\caption{Example of fitting, as in Figure~\ref{fig:example1}, but for
$n=10$. In this case the closest spatial leaks ($\delta m = \pm 2$,
$\delta\ell=0$) are barely resolved, and blends with the main peak in the
$m$-averaged spectrum.
         \label{fig:example3}}
\end{figure}

\begin{figure}[!p]
\includegraphics{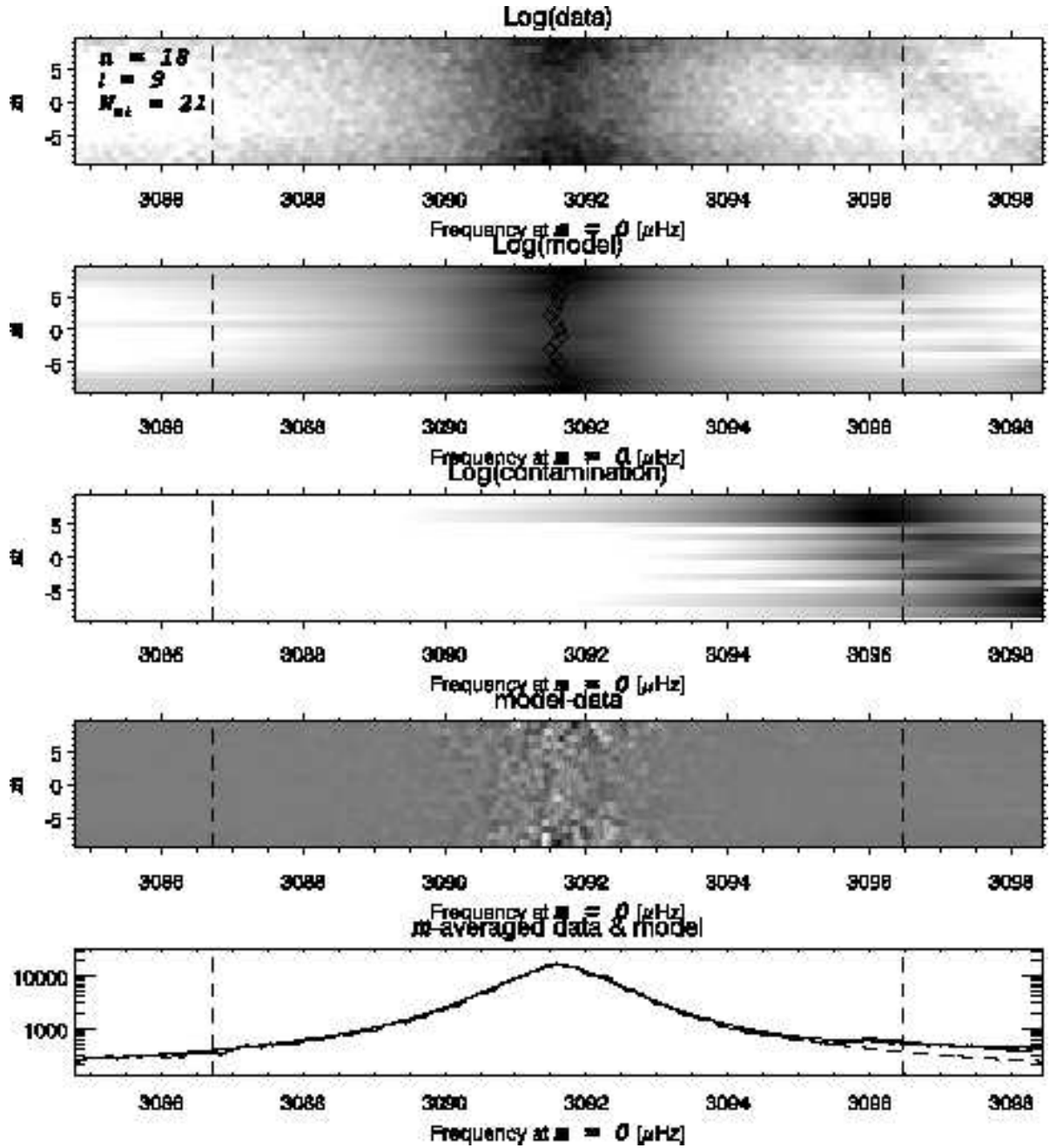}
\caption{Example of fitting, as in Figure~\ref{fig:example1}, but for $n=18$,
to illustrate the $n'=n\pm1$, $\ell'=\ell\mp3$ contamination. The
contamination -- illustrated by itself in an additional panel -- is in this
case relatively well separated from the main peak. The $m$-averaged model
without including this contamination is illustrated in the bottom panel by the
dashed line.  \label{fig:example4}}
\end{figure}

\begin{figure}[!p]
\includegraphics{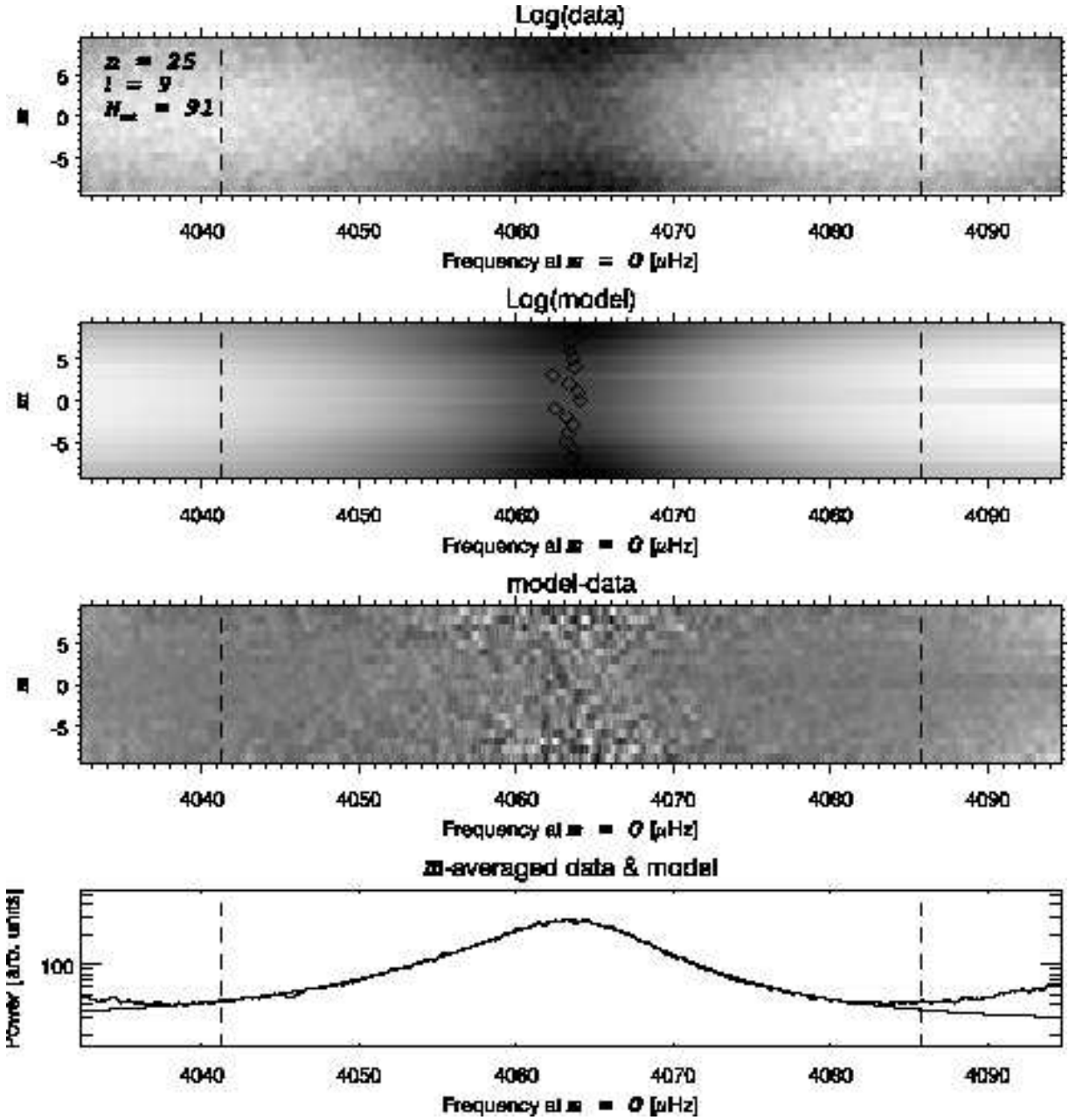}
\caption{Example of fitting, as in Figure~\ref{fig:example1}, but for
$n=25$, where the mode linewidth is very large and a large number of tapers
has been used to estimate the power spectrum ($N=91$). 
         \label{fig:example5}}
\end{figure}

  Figures~\ref{fig:example1} to \ref{fig:example5} illustrate the fitting of
the MDI 2088-day-long time series for $\ell=9$ and a selection of values of
$n$. It shows that for $n=4$ not all azimuthal orders emerge above the
background; it illustrates with $n=7$ the case where modes and leaks are well
resolved; it shows for $n=10$ a case where the leaks nearly merge with the
main peak; for $n=18$ it illustrates the $n'=n+1$, $\ell'= \ell-3$
contamination and finally for $n=25$ it presents an example of fitting a high
order mode where the optimal spectral estimator uses a large number of
multi-tapers.

\begin{figure}[!p]
\includegraphics{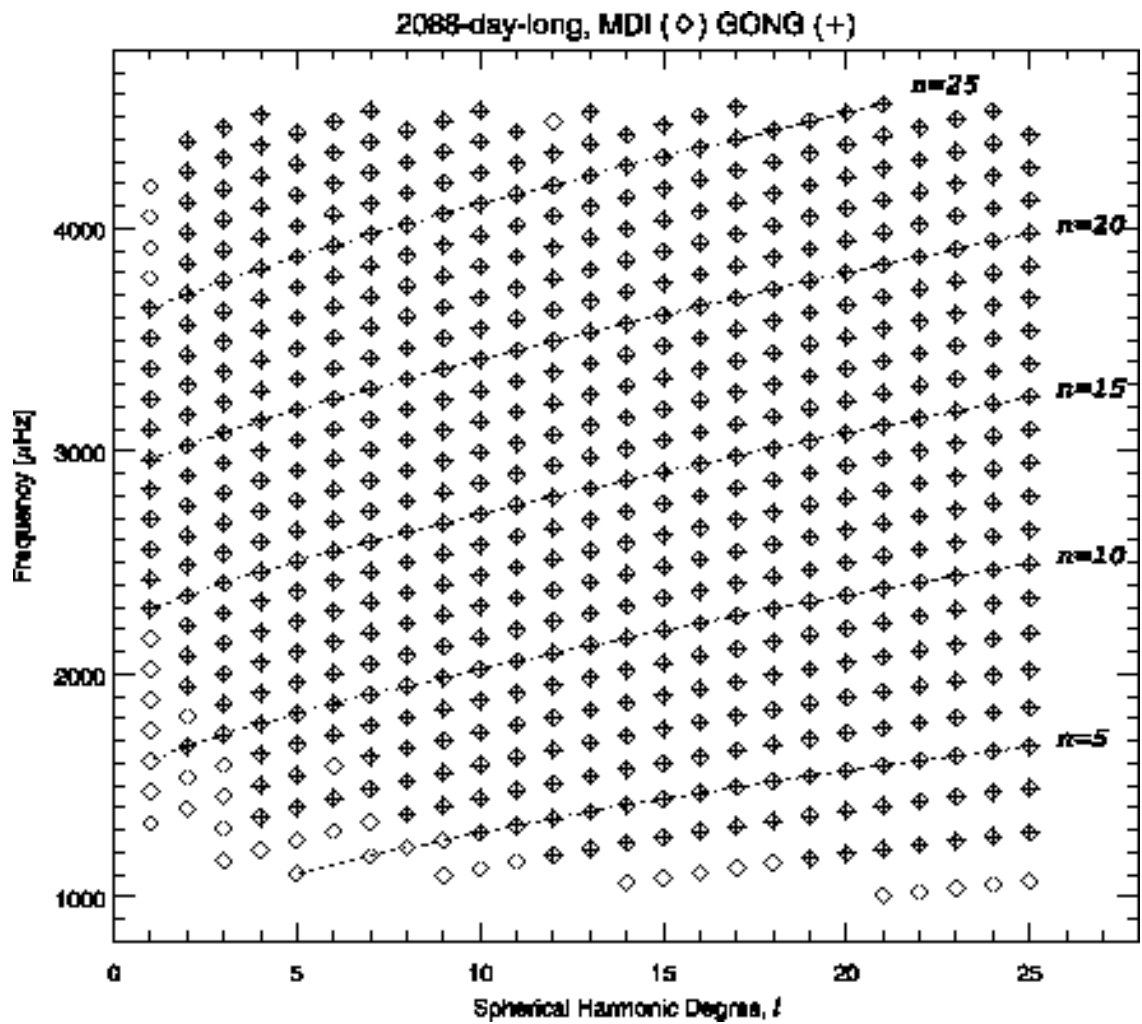}
\caption{The coverage, in a $\ell - \nu$ diagram, of the fitting range for the
2088-day-long time series, with diamonds for the MDI observations and crosses
for the GONG observations. The higher fill factor and lower signal-to-noise
ratio of the MDI time series allowed to extend the fitting towards lower
frequencies.  \label{fig:lnu}}
\end{figure}

  I have produced frequency tables of individual multiplets based on MDI and
GONG observations for 2088-day-long time series as well as five 728-day-long
time series, for $1 \le \ell \le 25$. The coverage in the $\ell - \nu$ diagram
is illustrated in Figure~\ref{fig:lnu} and corresponds to some 15,491 \&
14,883 $(n,\ell,m)$ multiplets or some 595 \& 554 $(n,\ell)$ singlets for MDI
and GONG 2088-day-long time series respectively.  The higher fill factor and
lower signal-to-noise ratio (SNR) of the MDI time series has allowed me to
push the fitting of that time series towards lower frequencies.

\begin{figure}[!p]
\includegraphics[scale=.975]{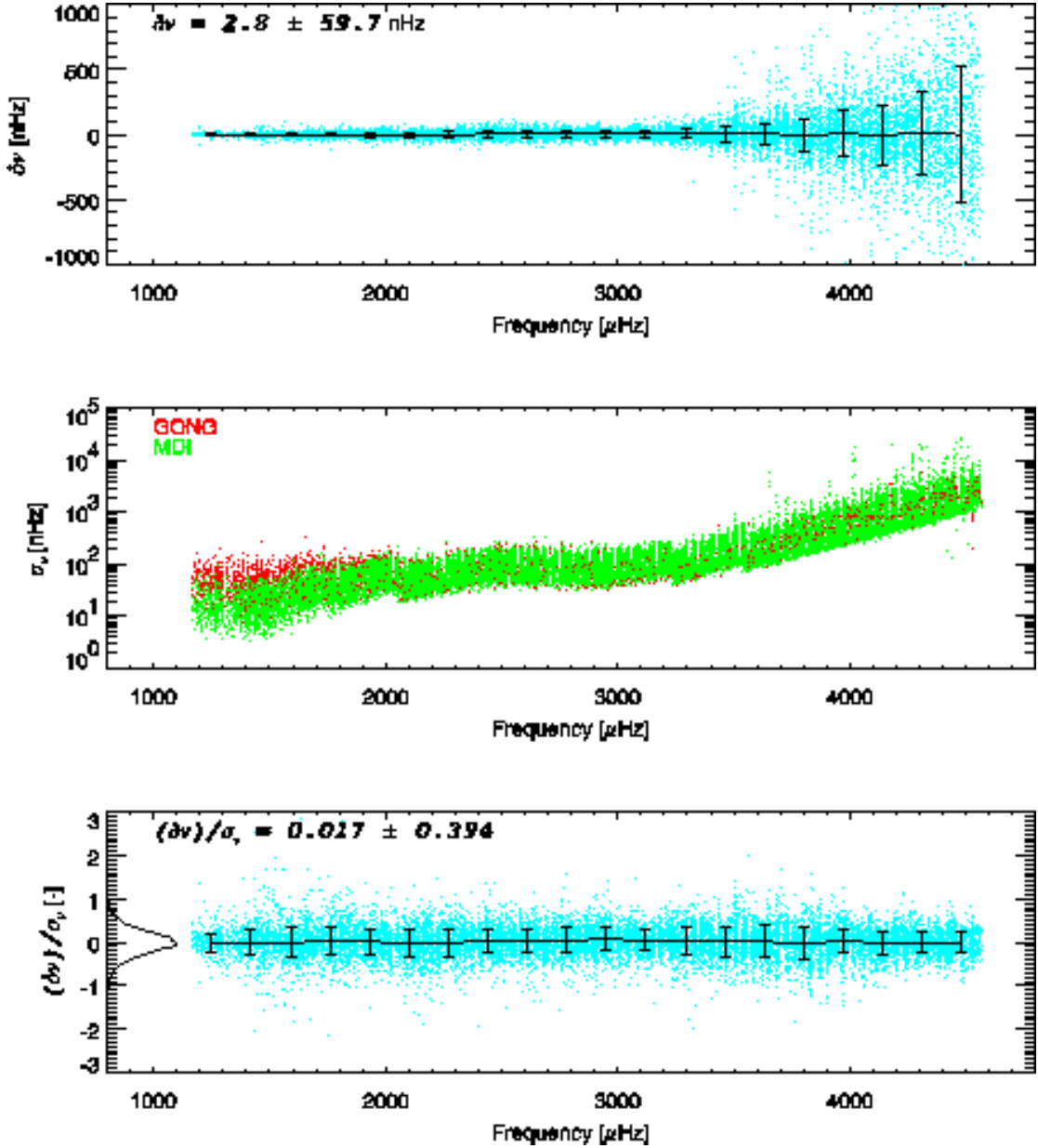}
\caption{Comparison of fitting results (multiplets) from the two 2088-day-long
co-eval MDI and GONG time series. Top panel shows frequency differences while
the bottom panel shows the frequency differences scaled to their uncertainties
(blue points); in both plots the solid line results from bining the points
over 20 equispaced frequency intervals, with the error bars representing the
standard deviation within the bin. A histogram of the scaled difference is
drawn in the bottom panel. The middle panel shows the frequency uncertainties.
\label{fig:compare1}}
\end{figure}

  Figure~\ref{fig:compare1} compares the results from fitting the two
2088-day-long co-eval time series, \ie\ MDI and GONG. The frequency differences
are small ($2.8 \pm 59.7$ nHz) and, as expected, increase with frequency --
since the accuracy of the fitting decreases as the mode linewidth
increases. The reduced frequency differences (${\delta\nu}/{\sigma_{\nu}}$)
are uniform and correspond to a nearly Gaussian distribution with a mean of
0.017 and a standard deviation of 0.394. The smaller than expected value of
the standard deviation suggests that my estimate of the uncertainties might be
too conservative. At low frequency, the error bars on the frequency for the
GONG observations are larger than the corresponding one for the MDI
observations again as the results of the difference in fill factor and SNR.

\begin{figure}[!p]
\includegraphics[scale=.975]{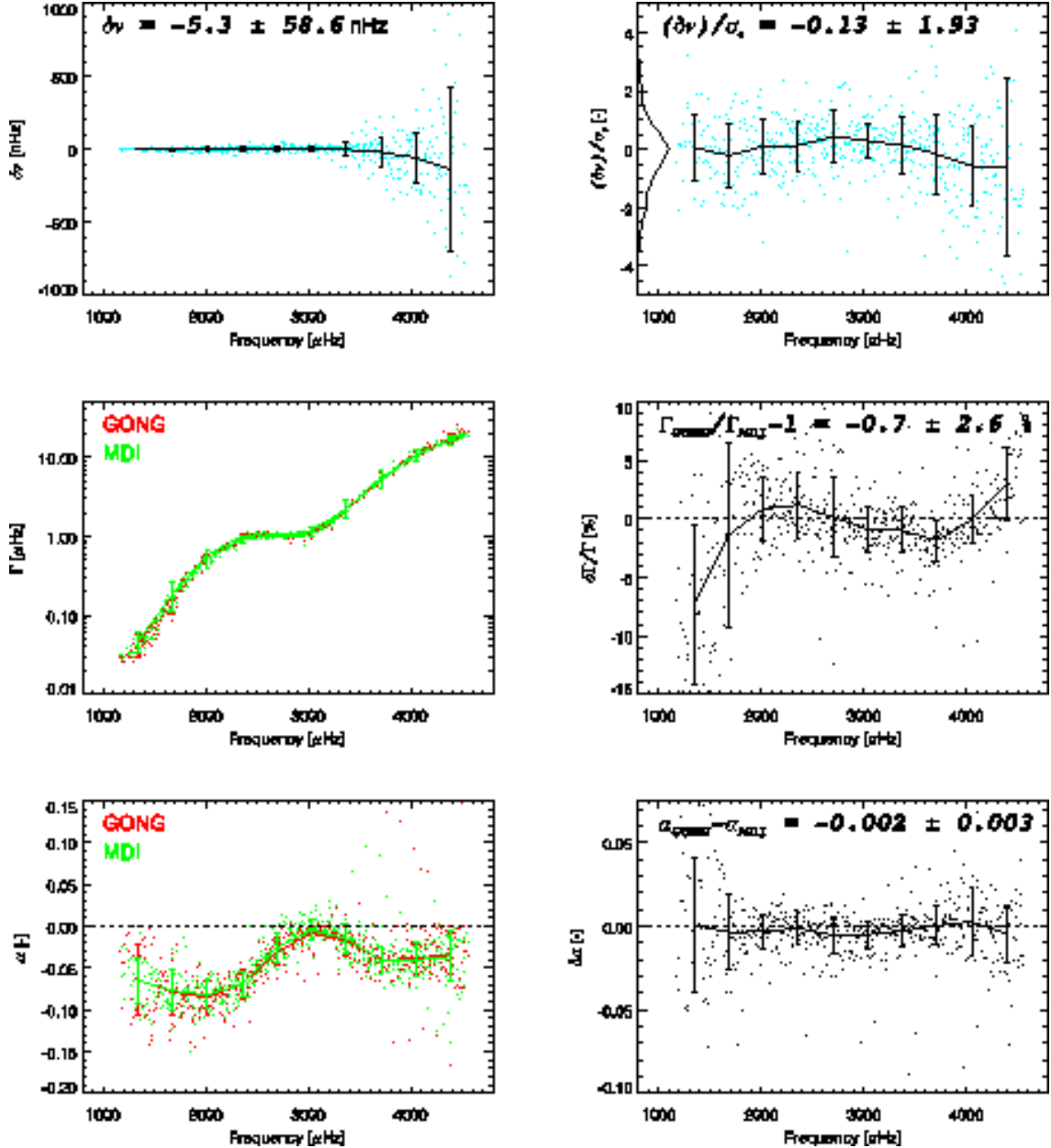}
\caption{Comparison of fitting results (singlets) from the two 2088-day-long
co-eval MDI and GONG time series (top panels). Middle panels compare the mode
linewidth ($\Gamma$) while the bottom panels compare the mode asymmetry
($\alpha$).  The solid lines correspond to bining individual points (shown as
dots) over 10 equispaced frequency intervals, with the error bars representing
the standard deviation within the bin.
\label{fig:compare2}}
\end{figure}

  Figure~\ref{fig:compare2} compares the mode linewidth ($\Gamma_{n,\ell}$)
and the asymmetry coefficients ($\alpha_{n,\ell}$), as well as the singlet
mode frequencies. The mode singlet was computed by fitting a Clebsch-Gordan
polynomial expansion \citep{Ritzwoller+Lavely:91} to the multiplets, with a
3$\sigma$ rejection. Mode linewidths and asymmetries between both
data sets are nearly identical.

  Comparisons for the five 728-day-long segments show similar results, and are
quantitatively summarized in Table~\ref{tab:comparison}. The larger average
frequency difference seen for segment no.\ 2 results most likely from the data
gap at the end of that time series present in the MDI
observations\footnote{This gap results from the loss of contact with the SOHO
spacecraft.}. Since the gap is at the end of that time series, the results for
MDI and GONG observations do not correspond to the same mean solar activity
level.

\begin{table}[!t]
\caption{Quantitative Comparison, MDI versus GONG. \label{tab:comparison}}
\begin{tabular}{||r|r@{$\,\pm\,$}l|r@{$\,\pm\,$}l|r@{$\,\pm\,$}l|r@{$\,\pm\,$}l||}
                          & \multicolumn{2}{c|}{$\delta\nu$}
                          & \multicolumn{2}{c|}{${\delta\nu}/{\sigma_\nu}$}
                          & \multicolumn{2}{c|}{${\delta\Gamma}/{\Gamma}$}
                          & \multicolumn{2}{c||}{$\delta\alpha$} \\
                          & \multicolumn{2}{c|}{[nHz]}
                          & \multicolumn{2}{c|}{~}
                          &\multicolumn{2}{c|}{[\%]}
                          & \multicolumn{2}{c||}{~}               \\\hline
2088-day-long             & $ 2.8$ & $59.7 $ & $ 0.017$ & $0.394$ & $-0.6$ & $2.7$ & $-0.001$ & $0.003$ \\
728-day-long seg.\ no.\ 1 & $-2.9$ & $93.5 $ & $-0.009$ & $0.389$ & $ 1.5$ & $3.1$ & $ 0.002$ & $0.005$ \\  
                 {no.\ 2} & $25.9$ & $118.1$ & $ 0.124$ & $0.569$ & $-1.5$ & $3.6$ & $-0.002$ & $0.003$ \\
                 {no.\ 3} & $-5.7$ & $113.4$ & $-0.035$ & $0.660$ & $-1.8$ & $1.3$ & $-0.001$ & $0.003$ \\
                 {no.\ 4} & $ 2.8$ & $85.8 $ & $ 0.013$ & $0.791$ & $ 0.2$ & $1.1$ & $-0.002$ & $0.004$ \\
                 {no.\ 5} & $ 5.4$ & $100.0$ & $ 0.023$ & $0.392$ & $-0.1$ & $2.0$ & $-0.003$ & $0.006$ \\
\hline
\end{tabular}

\end{table}


\subsection{Comparisons with Previous Estimates}

\subsubsection{MDI Observations}

\begin{figure}[!p]
\includegraphics[scale=.9]{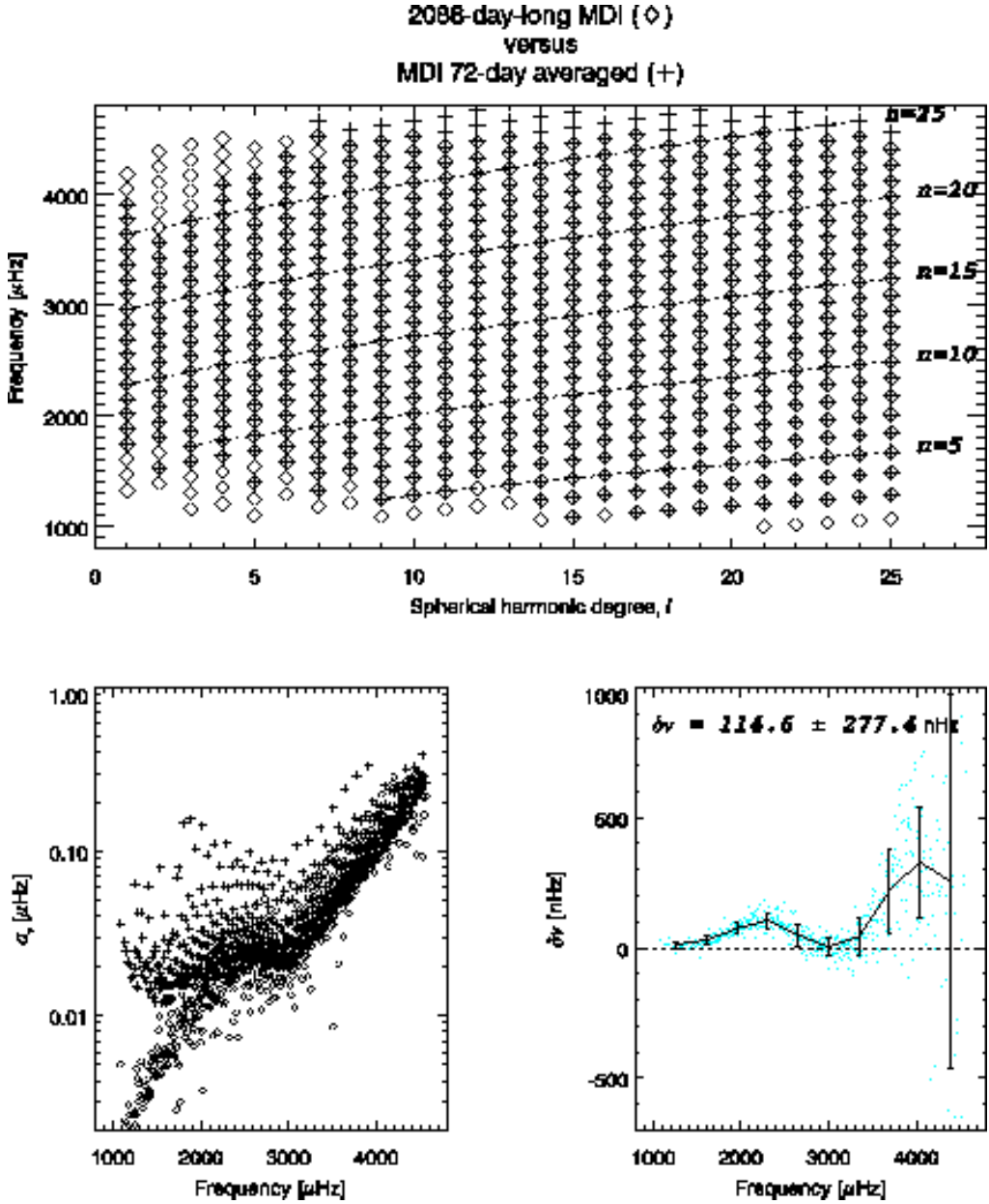}
\caption{Comparison between singlets resulting from fitting the 2088-day-long
MDI time series (diamonds) and the corresponding MDI average values computed
from 27 tables resulting from fitting 72-day-long times series. Top panel
shows the respective coverage in an $\ell$ -- $\nu$ diagram. The lower left
panel compares the frequency uncertainties, while the lower right shows the
frequency differences (dots) and these differences binned over 10 equispaced
frequency bins -- the error bars represent the standard deviation inside each
bin.
\label{fig:cmp_w_mdi1}}
\end{figure}

  Figure~\ref{fig:cmp_w_mdi1} compares results from my fit to the MDI
2088-day-long time series to average values computed from 27 
tables resulting from fitting 72-day-long times series that covers the same
time span\footnote{The missing 2 tables (since $2088=29\times72$)
correspond to the time interval when contact with the SOHO spacecraft was
lost.} and are routinely computed by the MDI team \citep{Schou:99}. The
averaging was weighted by the uncertainty -- to reflect the relative fill
factor of each period. The comparison is for singlets, since the MDI fitting
procedure uses a polynomial expansion in $m$.

  The top panel illustrates the region of the $\ell$ -- $\nu$ diagram where
fitting a very long time series allows for detecting additional low order
modes. The comparison of the resulting fitting uncertainty (lower right) shows
-- as expected -- that at high frequency the uncertainty is dominated by the
mode linewidth while at low frequency by the length of the time series.

  This figure also shows a systematic difference between frequencies, with a
specific frequency dependence. One obvious reason for this difference is the
fact that Schou's fitting uses a symmetric profile while I am fitting an
asymmetric one. 

\begin{figure}[!p]
\includegraphics{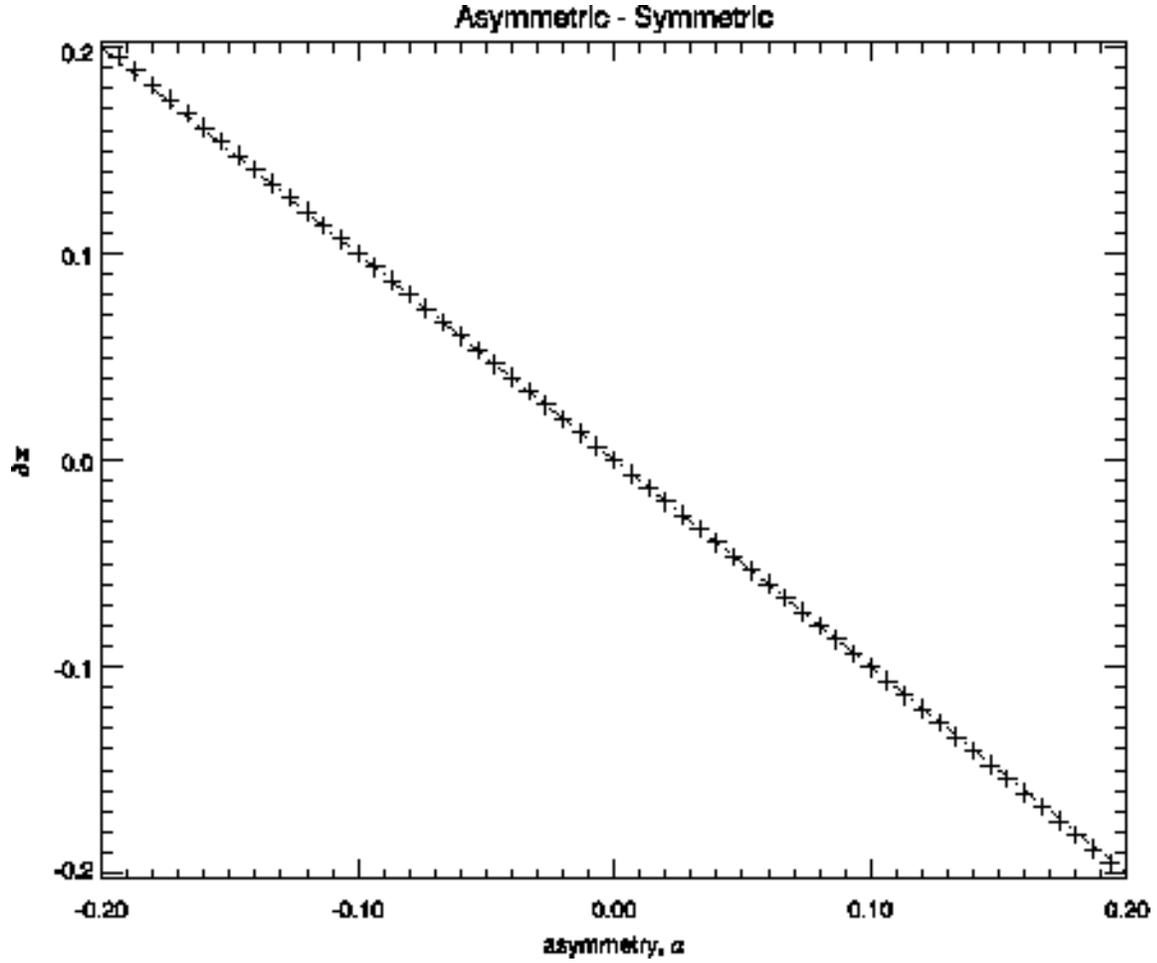}
\caption{Offset resulting from fitting a symmetric profile to an asymmetric
peak, as a function of the asymmetry coefficient, $\alpha$, for a fixed FWHM.
The predicted offset (crosses) scales nearly lineary with the asymmetry
coefficient (dash line). 
\label{fig:sym_vs_asym}}
\end{figure}

  To estimate the effect of fitting a symmetric profile to an asymmetric peak
I have computed a grid of isolated asymmetric profiles and fitted them with
symmetric ones. The resulting offset in frequency varies nearly linearly with
the asymmetry coefficient as defined in my parameterization\footnote{This
parameterization is equivalent to the one defined by Equation~4 of
\citet{Nigam+Kosovichev:1998}.  After some rudimentary algebra, one can
identify their asymmetry coefficient, $B$, to the one I use, $\alpha$, namely
$\frac{\alpha}{2} = B\,(1-\frac{\alpha^2}{2})$.}, for a given FWHM, as shown
in Figure~\ref{fig:sym_vs_asym}.  The systematic error introduced by fitting
an asymmetric peak with a symetric profile is thus given by
\begin{equation}
  \nu_{\rm asymmetric} - \nu_{\rm symmetric} = - \alpha\,\Gamma/2
  \label{eq:dnuasym}
\end{equation}

\begin{figure}[!p]
\includegraphics[scale=.95]{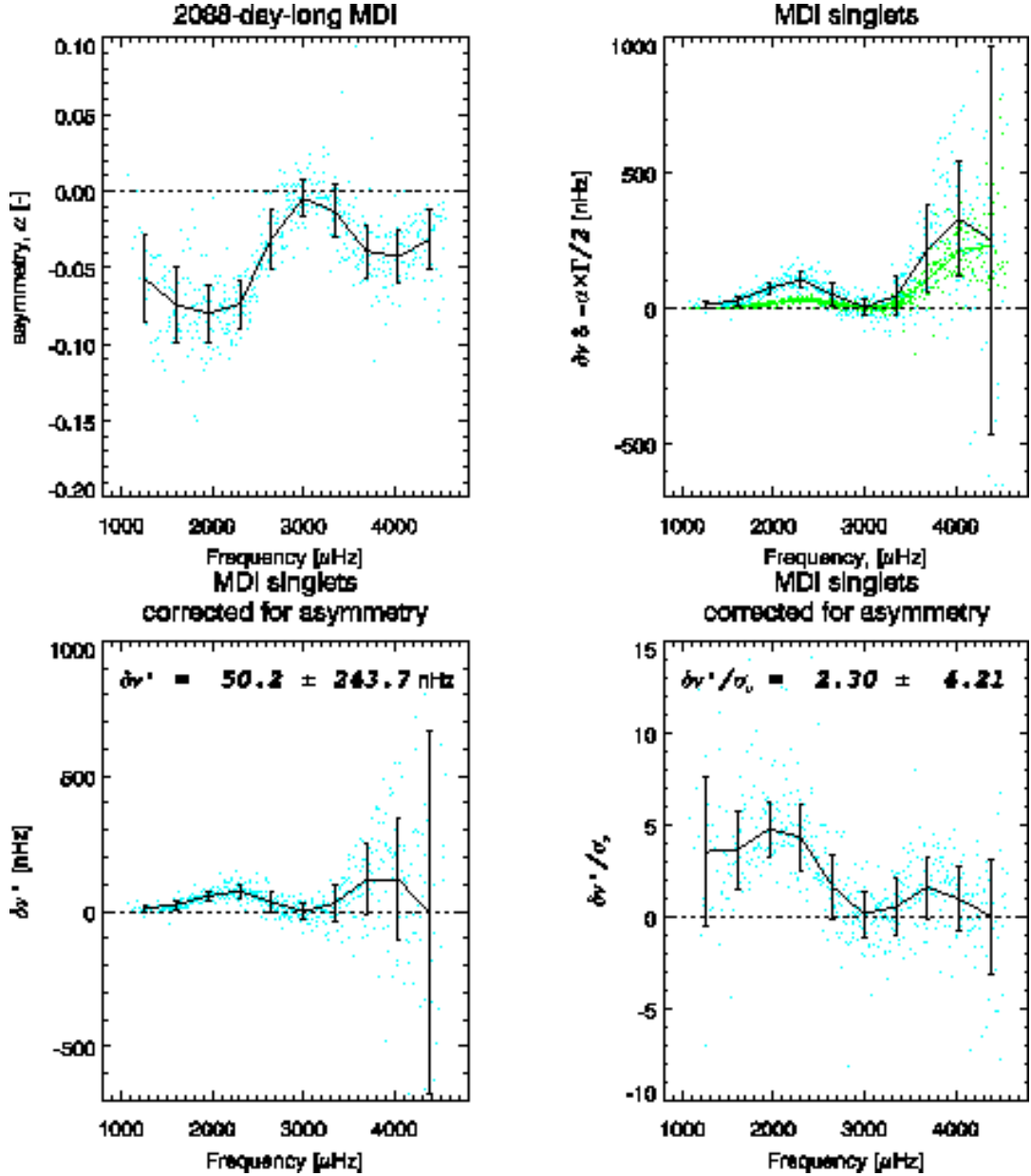}
\caption{Observed asymmetry (top left) and estimate of systematic frequency
offsets due to fitting a symmetric profile to an asymmetric peak (green points
in top right panel) compared to observed differences (blue points and solid
line in top right panel), for MDI values. The bottom panels show residual
differences after correcting for the effect of not including the mode
asymmetry. The solid lines represent the values (blue dots) binned over 10
equispaced frequency bins. The standard deviation inside each bin is
represented by the error bars.
\label{fig:cmp_w_mdi2}}
\end{figure}

  Roughly half of the frequency differences seen in
Figure~\ref{fig:cmp_w_mdi1} can be explained by this model, as shown
Figure~\ref{fig:cmp_w_mdi2}. The residual differences after correcting for the
asymmetry using Eq.~\ref{eq:dnuasym}, are marginaly significant, \ie\ at the
4$\sigma$ level, and still show a systematic trend with frequency.

\begin{figure}[!p]
\includegraphics{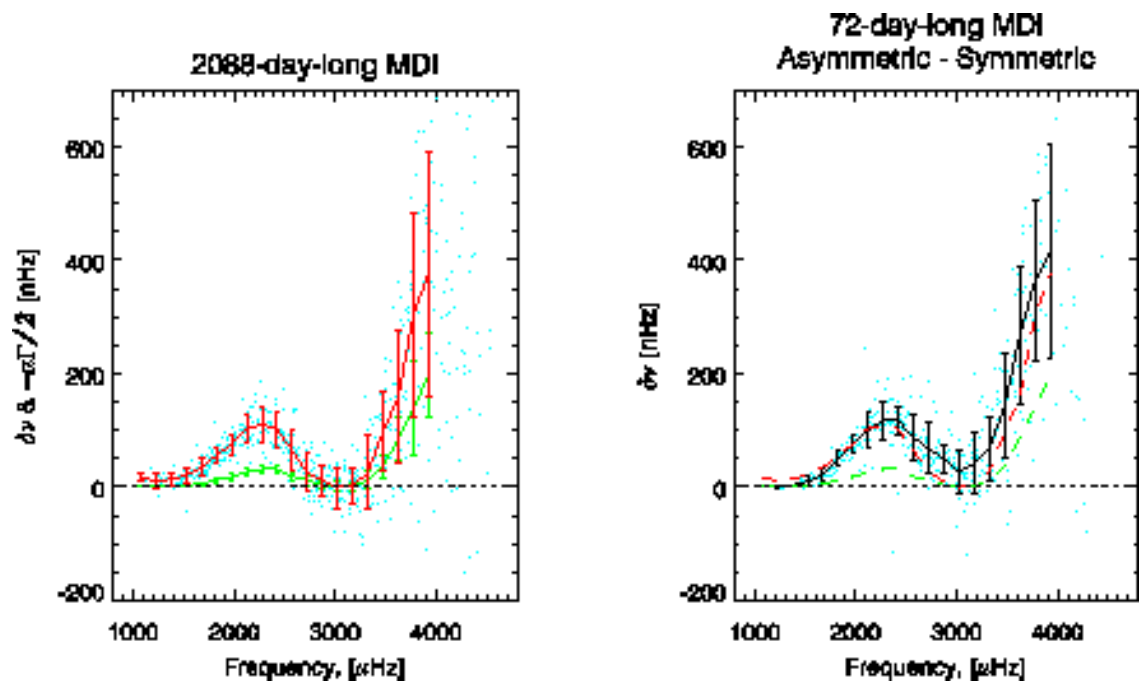}
\caption{(Left) Comparison of observed differences between fitting the
2088-day-long MDI time series and the corresponding MDI average values
computed from 27 tables resulting from fitting 72-day-long times series (raw
values as blue dots and binned values shown as the red line) with the
predicted differences resulting from a simple model (green points and line) of
fitting an isolated symmetric profile to an asymmetric one. (Right) Frequency
differences resulting from symmetric and asymmteric fits carried out by Schou
using a 72-day-long MDI time series (individual values shown as dots, solid
line represents binned values) compared to the frequency differences shown in
the left panel (red line, binned values) and the simple model prediction
(green line, binned values).\label{fig:cmp_sym_vs_asym}}
\end{figure}

  Back in 1991, Schou (private communication) fitted one 72-day-long MDI time
series\footnote{It is the 72-day-long time series starting at mission day no.\
2368, namely on June 27th, 1999 at 0 UT.} using an asymmetric profile as well
as a symmetric one.  The comparison between his symmetric and asymmetric
fitting for that time series and for modes up to $\ell = 25$ is shown in
Figure~\ref{fig:cmp_sym_vs_asym}. The frequency differences compare very well
with the systematic differences seen in Figure~\ref{fig:cmp_w_mdi1} and are
nearly twice as large as the values predicted by Eq.~\ref{eq:dnuasym}.  While
one could be tempted to rescale the prediction of Eq.~\ref{eq:dnuasym} by some
{\em ad hoc} factor, I must point out that not only there is no rationale for
this -- unless either the leakage matrix or the horizontal to vertical ratio
or both are substantially wrong -- but also that a single factor would not
match the observed differences (such factor would have to be quite different
for frequencies above and below 3 mHz).

\begin{figure}[!p]
\includegraphics[scale=.89]{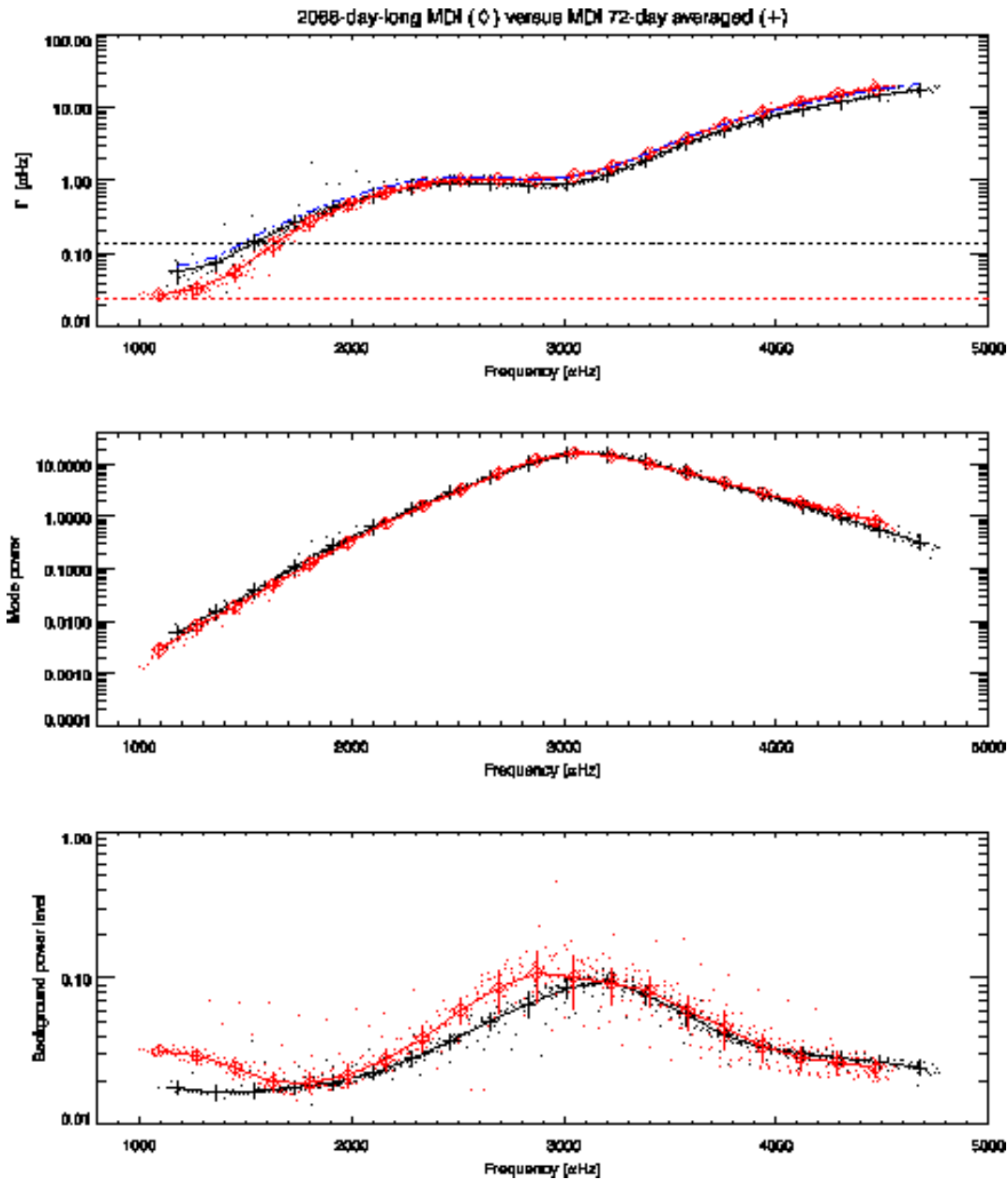}
\caption{Comparison between results from fitting the 2088-day-long MDI time
series (red dots \& diamonds) and the corresponding MDI average values
computed from 27 tables resulting from fitting 72-day-long times series (black
dots \& crosses).  The top panel compares linewidths, the horizontal lines
correspond to the respective time series frequency resolution, while the blue
curve illustrates the factor 1.2 between the two sets. The middle panel
compares the mode power ($A\times\Gamma$). The bottom panel compares the
background power level.
\label{fig:cmp_w_mdi3}}
\end{figure}

  Comparison of mode linewidths, amplitudes and background levels are shown in
Figure~\ref{fig:cmp_w_mdi3}.  The mode linewidth comparison shows a small
systematic discrepancy (\ie, a factor of 1.2) above 2.5 mHz, while below 2.5
mHz it indicates that the MDI estimates are too large -- most likely as a
result of the resolution limit of the 72-day-long time series -- despite being
corrected for the intrinsic frequency resolution of the time series. The mode
power and background level estimates compare rather well.

\subsubsection{GONG Observations}

\begin{figure}[!p]
\includegraphics[scale=.82]{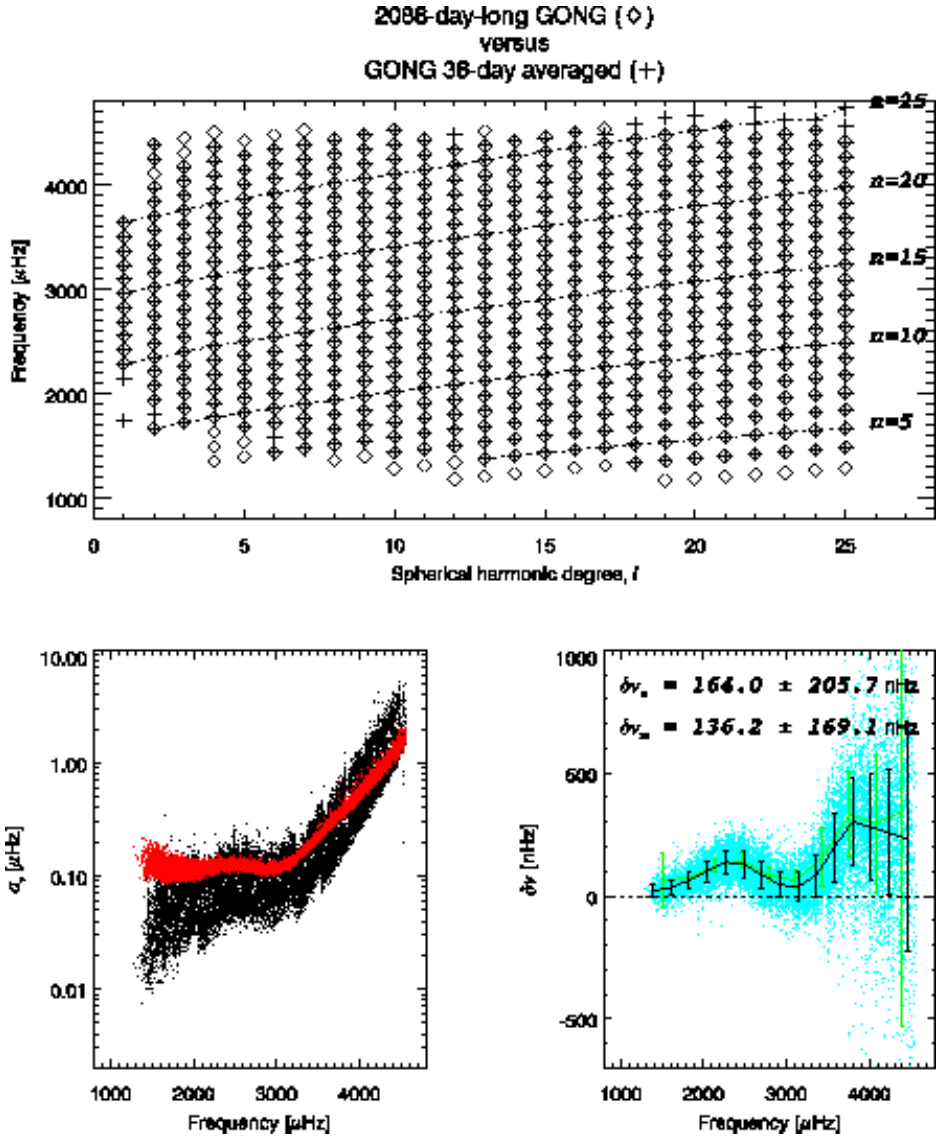}
\caption{Comparison between results from fitting the 2088-day-long GONG time
series (diamonds or black dots) and the corresponding GONG average values
computed from 58 tables resulting from fitting 36-day-long times series
(crosses or red dots). Top panel shows the respective coverage in an $\ell$ --
$\nu$ diagram, when computing averages only if the multiplet is present in at
least 10 of the 58 tables.  The lower left panel compares the frequency
uncertainties for the multiplets, while the lower right panel shows the
frequency differences (dots, multiplets) and the differences (singlets and
multiplets) binned over equispaced frequency bins -- the error bars represent
the standard deviation inside each bin. The binned singlets differences is
shown as the green curve. 
\label{fig:cmp_w_gong1}}
\end{figure}

  Figure~\ref{fig:cmp_w_gong1} compares results from my fit to the GONG
2088-day-long time series to average values based on 58 tables resulting from
fitting 36-day-long times series that covers the same time span and are
routinely computed by the GONG project \citep{HillEtAl:96}. The averaging was
weighted by the uncertainty -- to reflect the relative fill factor of each
period. Singlets were then computed -- by fitting a Clebsch-Gordan polynomial
expansion to the multiplets, with a 3$\sigma$ rejection of outliers -- to
produce comparisons similar to the case of the MDI observations.

\begin{figure}[!p]
\includegraphics{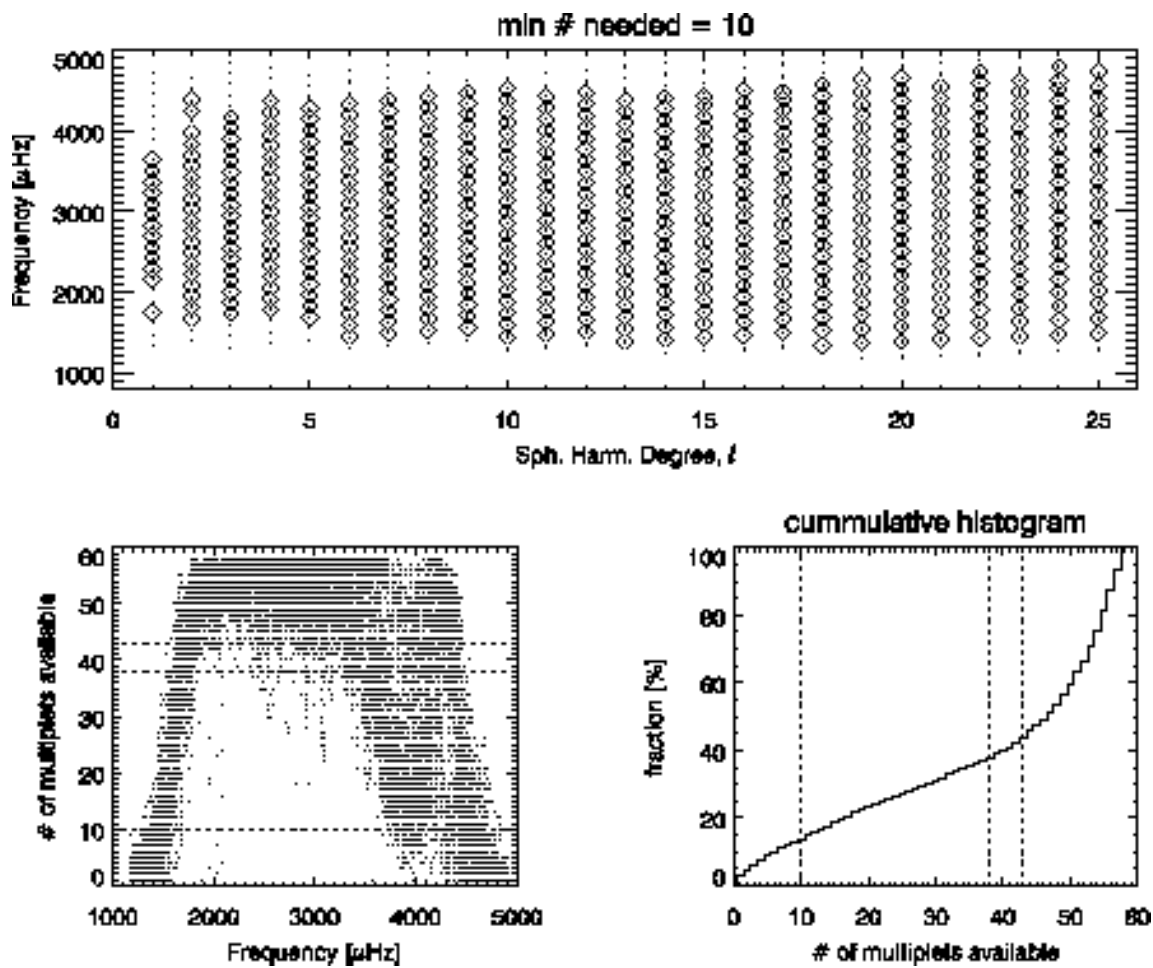}
\caption{This figure illustrates a consistency problem present in the GONG
frequency tables. The top panel shows in an $\ell$--$\nu$ diagram
averaged multiplets (dots) and the resulting average singlets when only
considering multiplets that are present in at least 10 of the 58 tables. The
lower left panel shows the number of times each multiplet is present -- as a
function of its frequency. The lower right panel shows the cumulative
histogram of the number of times a multiplet is present. Dashed lines are
draw at 10 (the value I ended up using) and at 2/3 and 3/4 of 58. When using
10 as a threshold only some 15\% of the mode set has to be dropped, while if I
would use a more conservative threshold nearly half of the mode set would be
ignored. Note also how using only such a threshold can skew singlet estimates
at the edge of the covered $\ell$--$\nu$ diagram.
\label{fig:gong_incons}}
\end{figure}

  Unfortunately the GONG fitting methodology does not produce consistent
frequency tables: for a substantial number of modes the same multiplets are
not fitted every time, as illustrated in Figure~\ref{fig:gong_incons}. The
averages used in the comparison in Figure~\ref{fig:cmp_w_gong1} correspond to
the case where the multiplet is measured at least 10 times out of 58. Using a
higher threshold -- to obtain better temporal averages -- would reduce
substantially the resulting frequency set.

  The lower SNR of the GONG observations at low frequency does not allow to
push mode fitting down to orders as low as for the MDI observations. But using
a long time series still allowed me to fit low order modes that are rarely
(less than 10 out of 58 times) fitted by the GONG project.

\begin{figure}
\includegraphics[scale=.9]{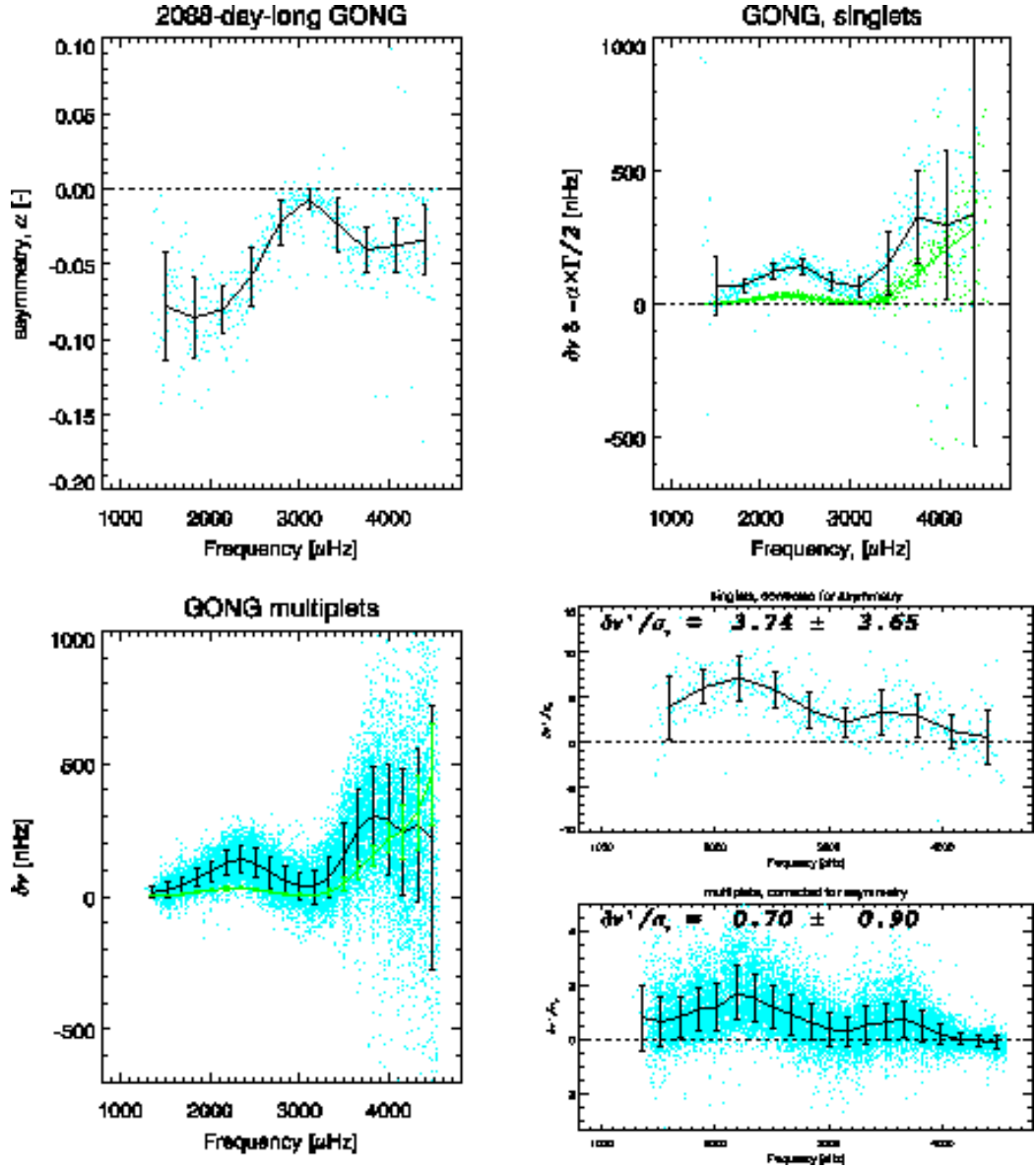}
\caption{Observed asymmetry (top left) and estimate of systematic frequency
offsets due to fitting a symmetric profile to an asymmetric peak (in green,
top right) compared to observed differences (blue dots and solid line) for
GONG singlets. The bottom left panel compares directly the multiplets (blue
dots and solid line) to the estimate of systematic offsets (green curve).
The bottom right panels show the reduced frequency differences, after
correcting for the effect of not including the mode asymmetry, for singlets
and multiplets.
\label{fig:cmp_w_gong2}}
\end{figure}

  The frequency comparison, presented in Figure~\ref{fig:cmp_w_gong2}, shows
systematic differences -- whether using singlets or multiplets\footnote{The
GONG project produces tables of multiplets.}. The GONG project also fits a
symmetric profile, hence some of the discrepency can be attributed to fitting
a symmetric profile to an asymmetric peak. The residual differences, after
correcting for the asymmetry using Eq.~\ref{eq:dnuasym}, are also shown in
Fig.~\ref{fig:cmp_w_gong2} for both singlets and multiplets. The corrected
singlets show marginally significant differences (at the 4$\sigma$ level) with
a systematic trend with frequency as for the MDI comparison (see
Fig.~\ref{fig:cmp_w_mdi2}). The corrected multiplets show the same trend with
frequency, but with less significance (0.8$\sigma$) simply because the
uncertainties on the multiplets are larger.

\begin{figure}[!p]
\includegraphics[scale=.9]{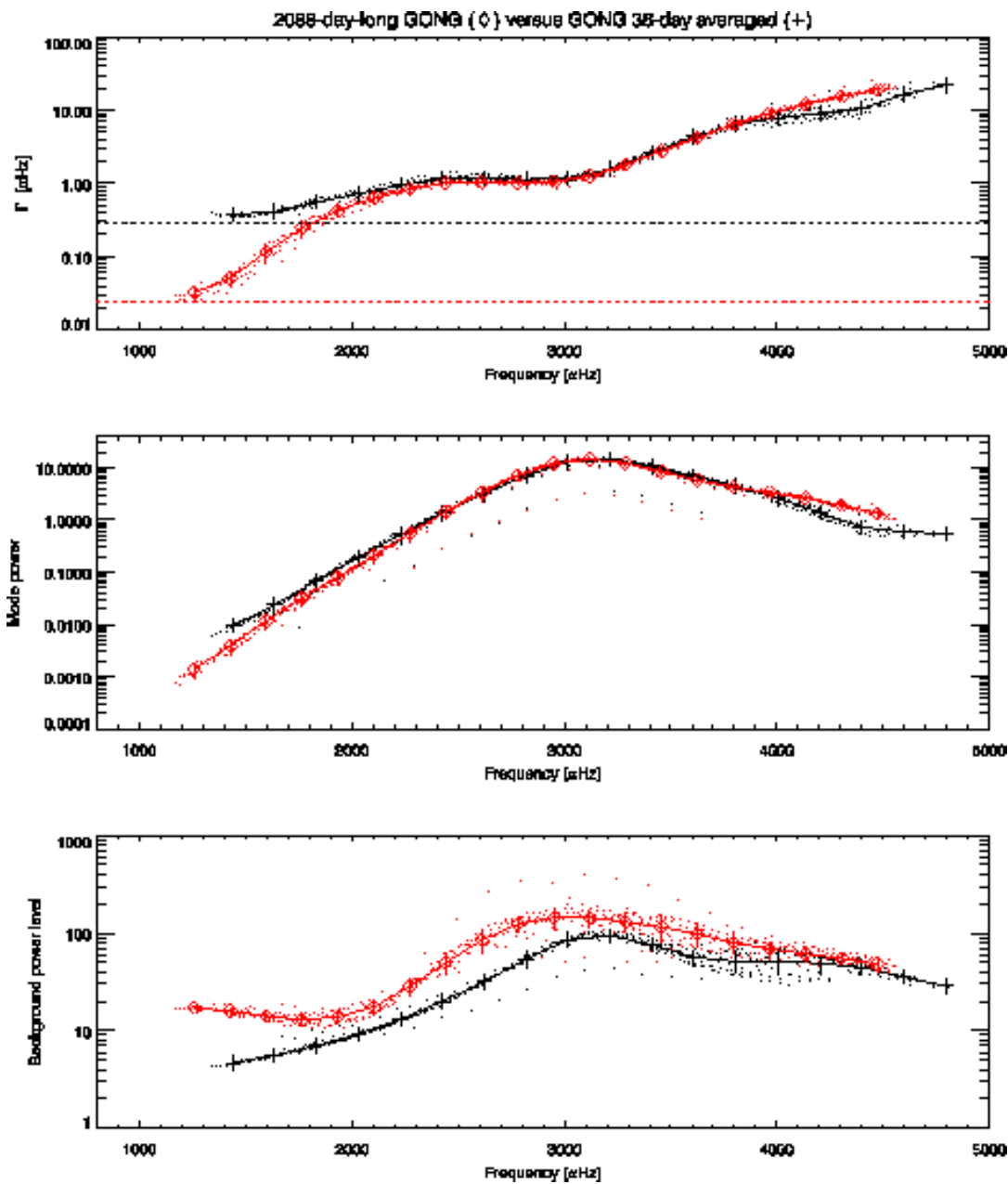}
\caption{Comparison between results from fitting the 2088-day-long GONG time
series (red dots \& diamonds) and the corresponding GONG average values
computed from 58 tables resulting from fitting 36-day-long times series (black
dots \& crosses).  The top panel compares linewidths, the horizontal lines
correspond to the respective time series frequency resolution. The middle
panel compares the mode power ($A\times\Gamma$). The bottom panel compares the
background power level.
\label{fig:cmp_w_gong3}}
\end{figure}

 Comparison of mode linewidth, presented in Figure~\ref{fig:cmp_w_gong3},
indicates that below 2.5 mHz the GONG values are dominated by the frequency
resolution of the time series -- not surprisingly, since by contrast to MDI's
estimates they are not corrected for that effect -- while agreeing rather well
above 2.5 mHz -- except for the dip between 4 and 4.5 mHz.  Mode power levels
compare rather well -- except at low and high frequencies -- while estimates
of the background level do not compare as well.

\subsubsection{MDI Observations - II}

\begin{figure}[!p]
\includegraphics{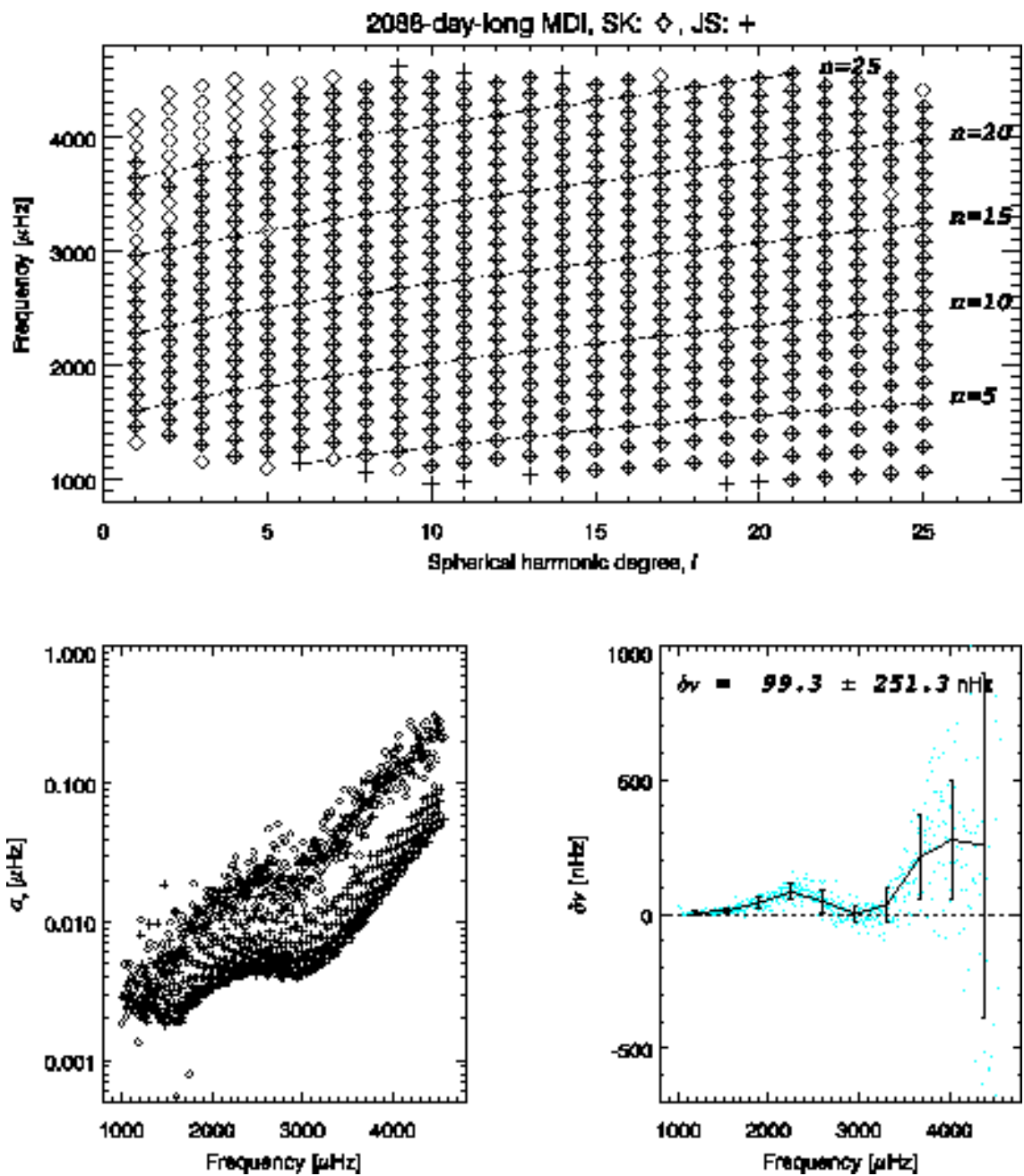}
\caption{Comparison of MDI singlets resulting from this work (crosses) and
from Schou's fitting to the same 2088-day-long time series
(diamonds). Coverage in the $\ell$ -- $\nu$ diagram is very similar. My
estimate of frequency uncertainties appears too conservative while the
frequency differences show a systematic pattern with frequency.
\label{fig:cmp_w_mdi2088d_1}}
\end{figure}

\begin{figure}[!p]
\includegraphics[scale=.95]{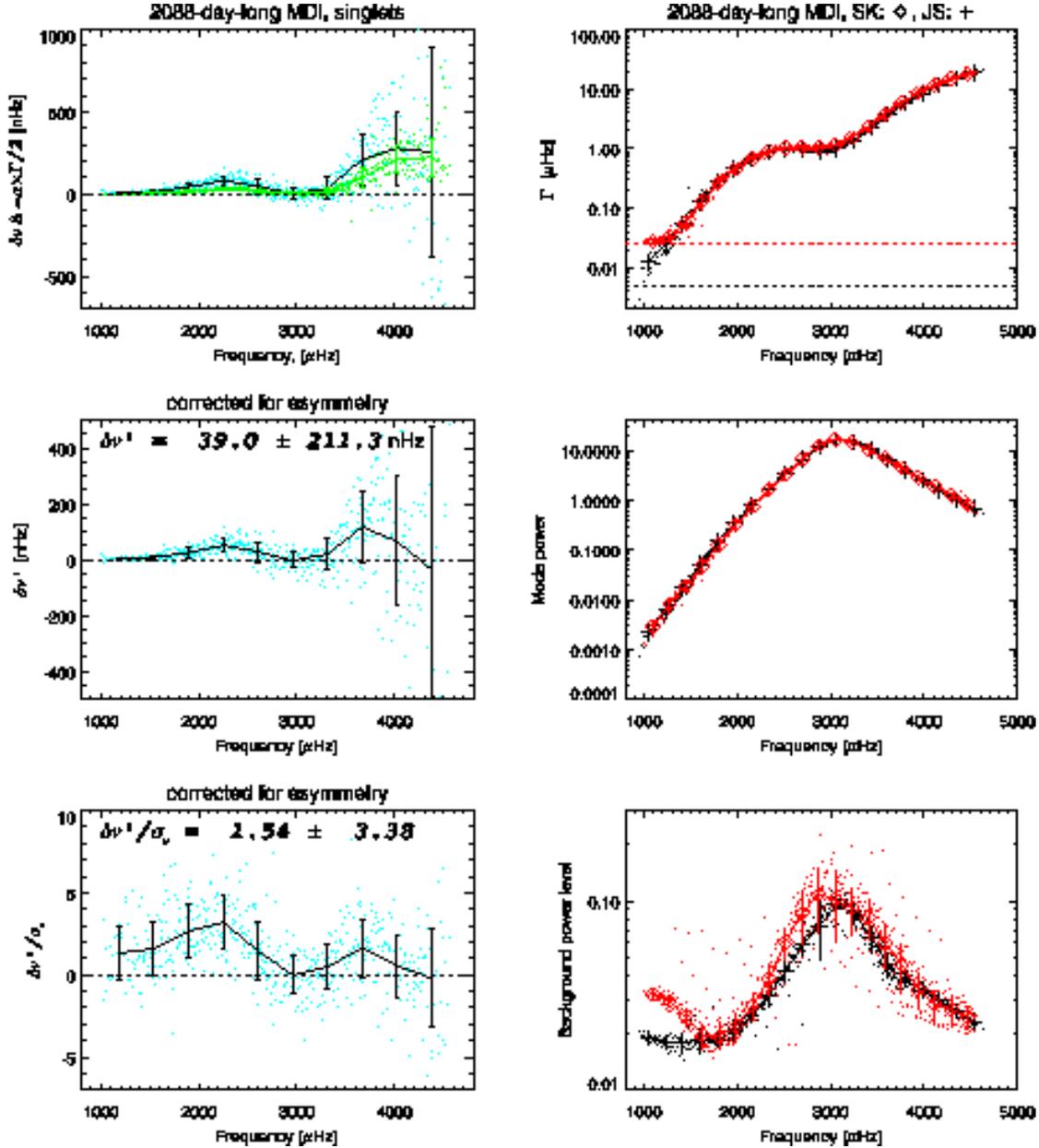}
\caption{Comparison of MDI singlets resulting from this work (crosses) and
from Schou's fitting to the same 2088-day-long time series (diamonds). The
frequency differences (blue dots and black curves for binned values) are not
fully explained by the effect of not including an asymmetry in the mode
fitting (green points, top left panel). Residual differences, after correcting
for the effect of not including the mode asymmetry, remain at the 3$\sigma$
level.  Mode linewidth and power as well as estimates of background level
compares remarkably well.
\label{fig:cmp_w_mdi2088d_2}}
\end{figure}

  Schou (private communication, 2004) has carried out a fitting of the
2088-day-long time series as well -- using the same methodology he uses for
fitting the 72-day-long time series. He has graciously provided me with the
results of that fit for a direct comparison that is shown in
Figures~\ref{fig:cmp_w_mdi2088d_1} and~\ref{fig:cmp_w_mdi2088d_2}.  Coverage
over the $\ell - \nu$ diagram is comparable, while the comparison of the
uncertainties on the frequency suggests -- as indicated earlier -- that my
estimates might be too conservative, by as much as a factor 3. The comparison
of the mode frequencies (singlets) shows again systematic differences, that
are in part explained by the effect of not including the mode asymmetry.

  The residual frequency differences after correcting for the asymmetry using
Eq.~\ref{eq:dnuasym} are at the 3$\sigma$ level and present a residual
systematic variation with frequency.
  The mode linewidths agree remarkably well (the discrepancy factor of $1.2$ is
no longer present) -- except at very low frequency where my uncorrected
estimates are biased by the frequency resolution. The mode power and
background level estimates compare rather well.

\subsection{Changes with Epoch}

\subsubsection{MDI \& GONG Observations}

\begin{figure}[!p]
\includegraphics[scale=.95]{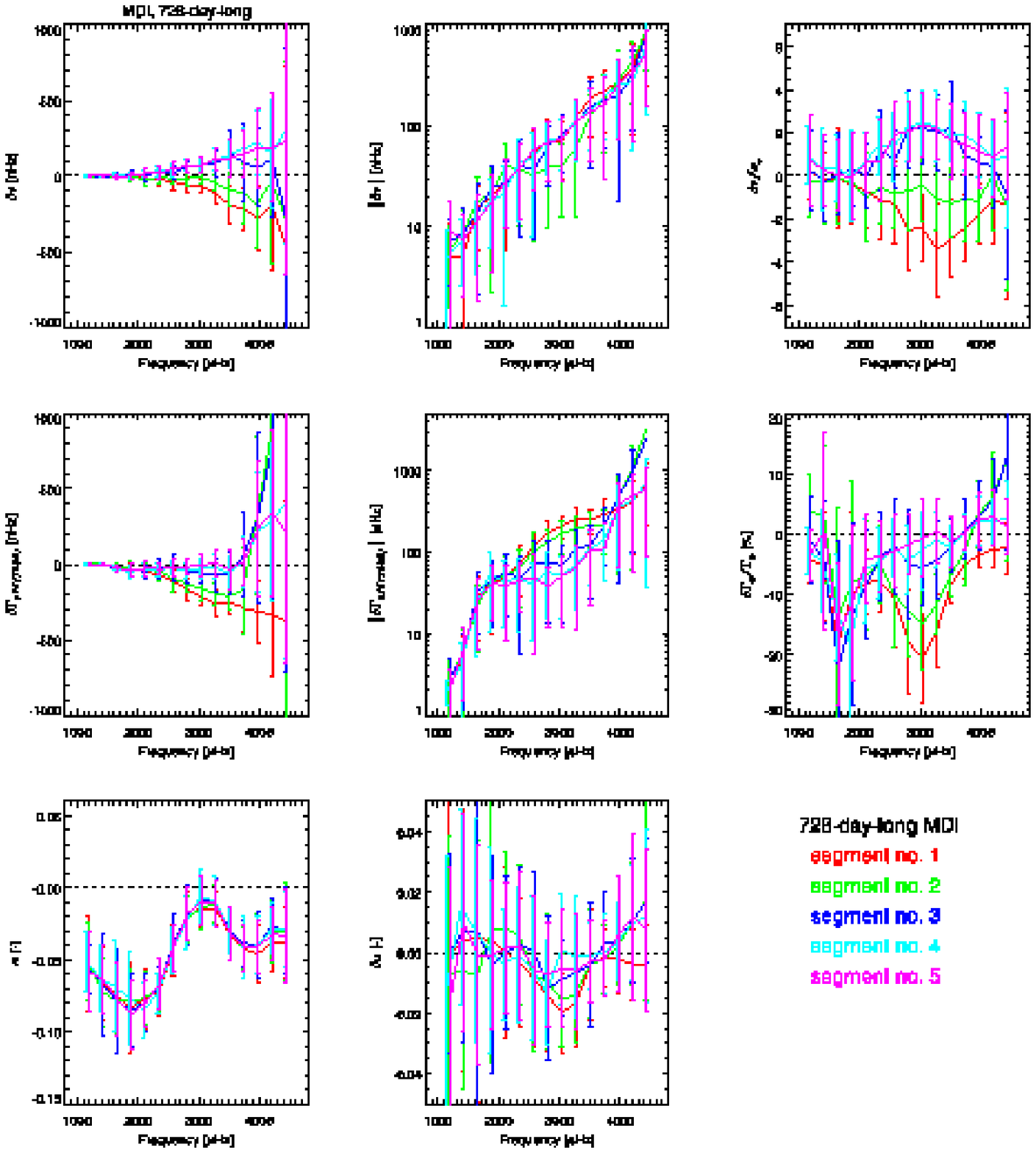}
\caption{Changes in singlet parameters, with respect to the values estimated
from the 2088-day-long time series, using MDI observations over the five
728-day-long segments.  Panels in the top row show change in frequency, panels
in the middle row show changes in linewidth, while the panels in the bottom
row compare estimates of asymmetry. Binned values, over equispaced interval in
frequency are shown, with the standard deviation within the bin indicated by
the error bars.  \label{fig:tc_mdi}}
\end{figure}

   Changes in frequency, linewidth and asymmetry, with respect to the values
estimated from the 2088-day-long time series, and using MDI observations over
the five 728-day-long segments are illustrated in Figure~\ref{fig:tc_mdi}.
This figure shows that the frequency changes with epoch scale with frequency
and range from 5 to 800 nHz, a now well established property that can be
explained by changes with activity levels concentrated near the solar surface.
The frequency changes at low degrees and very low frequency are very small and
barely significant. These changes are thus small enough to justify using very
long time series to fit these modes.

  Changes in linewidth with epoch also scale with frequency, with relative
changes as large as 20\% observed around 3 mHz. Changes at low frequency are
hard to measure since the measured width is dominated by the spectral
resolution.  The mode asymmetry shows no significant sign of change with time.

\begin{figure}[!p]
\includegraphics[scale=.95]{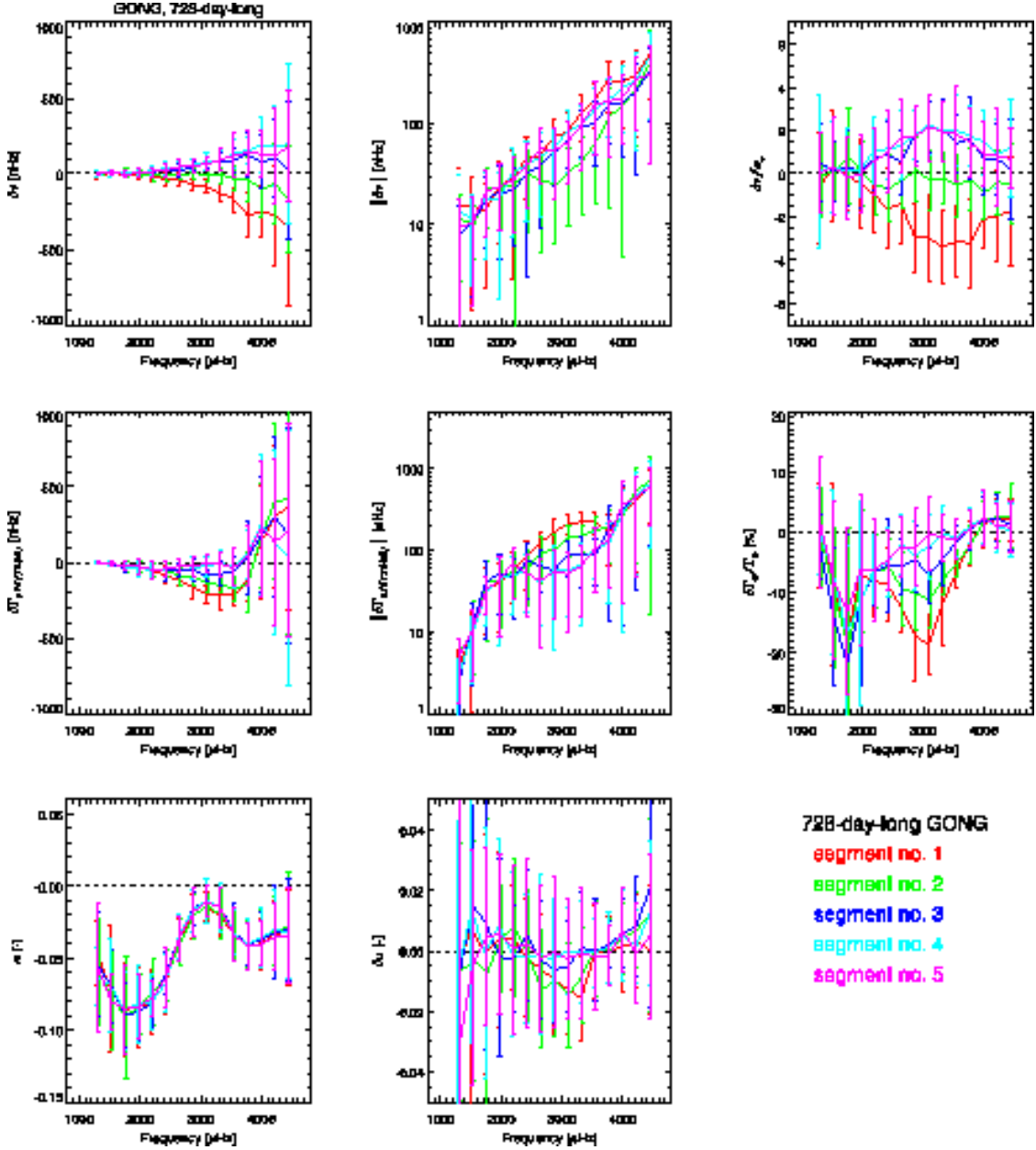}
\caption{Changes in singlet parameters, with respect to the values estimated
from the 2088-day-long time series, using GONG observations over the five
728-day-long segments.  Panels in the top row show change in frequency, panels
in the middle row show changes in linewidth, while the panels in the bottom
row compare estimates of asymmetry. Binned values, over equispaced interval in
frequency are shown, with the standard deviation within the bin indicated by
the error bars.  \label{fig:tc_gong}}
\end{figure}

  Changes in frequency, linewidth and asymmetry, with respect to the values
estimated from the 2088-day-long time series, and using GONG observations over
the five 728-day-long segments are shown in Figure~\ref{fig:tc_gong}. These
are very similar to the ones for the MDI data.

\subsubsection{Changes as a Function of Frequency and Azimuthal Order}

\begin{figure}[!p]
\hspace*{.5in}\includegraphics[scale=.97]{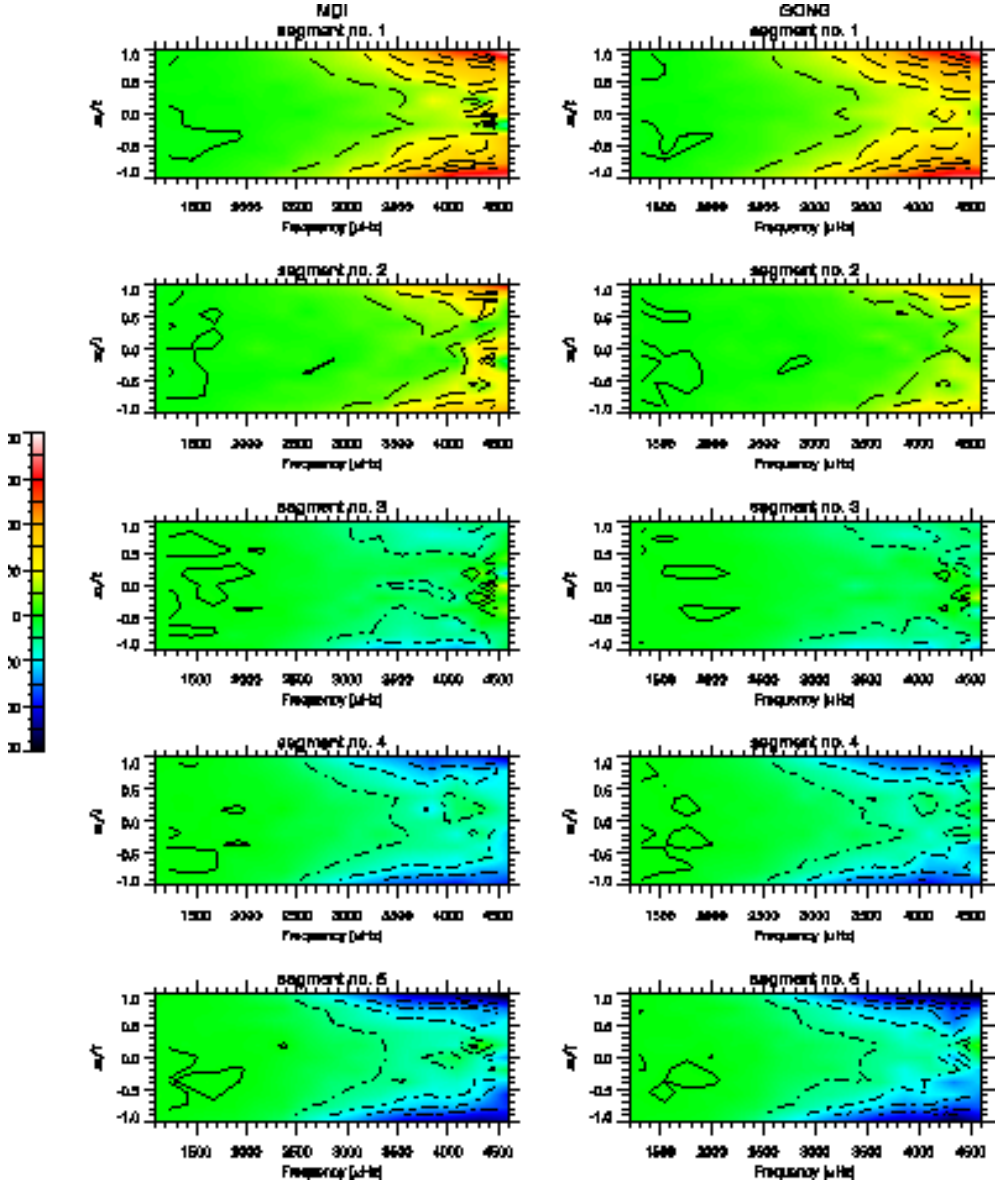}
\caption{Changes in frequency for each 728-day-long segment with respect to
the values estimated from the 2088-day-long time series based on multiplets
and plotted as a function of frequency and the ratio $m/\ell$, for MDI (left
column) and GONG (right column). The frequency differences were binned over a
grid equispaced in $\nu$ and $m/\ell$.  \label{fig:tc_multiplets}}
\end{figure}

\begin{figure}[!p]
\hspace*{1in}\includegraphics[scale=.84]{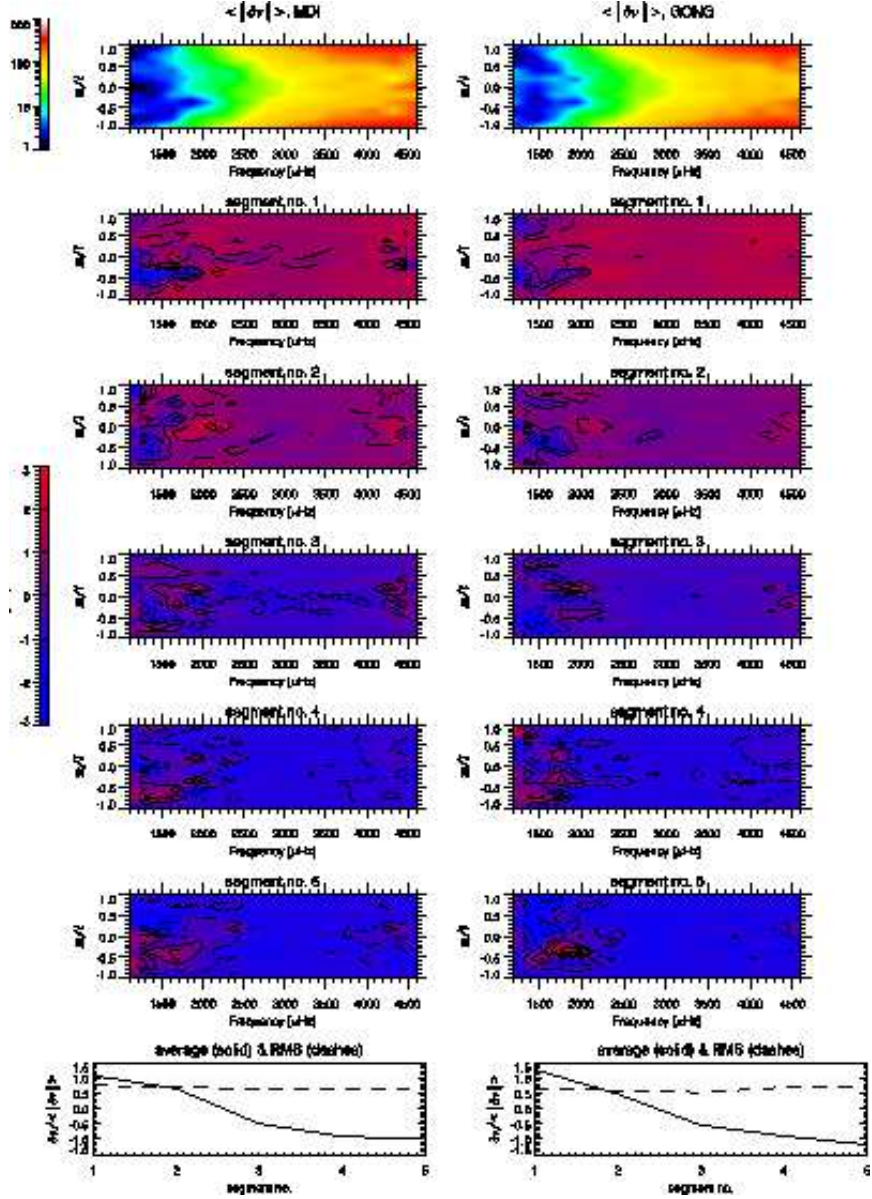}
\caption{Top panels: average of the absolute value of the changes in frequency
for each 728-day-long segment with respect to the values estimated from the
2088-day-long time series, as shown in Fig.~\ref{fig:tc_multiplets}. Middle 5
panels: frequency changes divided by the average of the absolute values, for
each 728-day-long segment. Bottom panels: average and RMS of the changes
scaled by the average of the absolute values as a function of segment
number. The RMS remains nearly constant while the average varies with time,
indicating that the changes in frequencies can be described by a fixed pattern
nearly constant in time scaled by a time dependant
factor. \label{fig:tc_multiplets_2}}
\end{figure}

  Changes in frequency, computed from multiplets rather than singlets, with
respect to epoch are shown for both data set in
Figure~\ref{fig:tc_multiplets}, as a function of frequency and azimuthal
order. The individual frequency differences were binned over an equispaced
grid in $\nu$ and $m/\ell$ to generate this figure. This figure shows that the
change in frequencies is concentrated near the sectoral modes and that it is
dominated by a pattern mostly symmetric in $m$ and nearly constant in time --
given a time dependant scaling factor. This is clearly illustrated in
Figure~\ref{fig:tc_multiplets_2} where the top panels show the average of the
absolute value of the binned frequency differences, while the midle panels
show how well the changes for each segment scale with that average. The bottom
panels show the variation of the mean scaled changes and the nearly constancy
of the RMS of these scaled changes.

\section{Conclusions}

  Fitting very-long time series has allowed me to push the precise
characterization of low degree modes to lower frequencies. The use of such a
very long time series is well justified for fitting of these modes since their
variation with activity remains small and comparable to the fitting
uncertainty itself. The resulting table of frequencies\footnote{Available at
{\tt ftp://cfa-ftp.harvard.edu/pub/sylvain/tables/}.} will allow us to further
improve inferences on the structure and dynamics of the solar deep interior.

  The methodology I have developed includes the most up-to-date procedural
elements: the use of an optimized sine multi-tapered power spectrum estimator;
the simultaneous fitting of all azimuthal orders -- each individually
parameterized; the use of the complete leakage matrix and the inclusion of an
asymmetric mode profile.

  The inter-comparison of the values resulting from my fit to the co-eval MDI
and GONG time series is very good. The comparison with equivalent values
produced by the MDI team and the GONG project is not as good. It is
complicated by the inclusion in my fit of the mode profile asymmetry. The
observed differences in mode frequencies are most likely dominated by the
effect of including or not the mode profile asymmetry.  Unfortunately a simple
model for this effect does not fully account for the discrepancies.

\acknowledgments
\section*{Acknowledgments}

  I am very grateful to J.~Schou for providing his leakage matrix coefficients
and results from his mode fitting, to R.~Howe and F.~Hill from providing the
GONG leakage matrix coefficients.

  The Solar Oscillations Investigation - Michelson Doppler Imager project on
SOHO is supported by NASA grant NAG5--8878 and NAG5--10483 at Stanford
University.  SOHO is a project of international cooperation between ESA and
NASA.

  This work utilizes data obtained by the Global Oscillation Network Group
(GONG) program, managed by the National Solar Observatory, which is operated
by AURA, Inc. under a cooperative agreement with the National Science
Foundation. The data were acquired by instruments operated by the Big Bear
Solar Observatory, High Altitude Observatory, Learmonth Solar Observatory,
Udaipur Solar Observatory, Instituto de Astrof\'{\i}sico de Canarias, and
Cerro Tololo Interamerican Observatory.

 SGK was supported by NASA grant NAG5--9819 \& NAG5--13501 and by NSF grant
ATM--0318390.

\clearpage

\clearpage

\clearpage

\end{document}